\documentstyle[preprint,aps,epsbox]{revtex}

\makeatletter

\def\lsim{\compoundrel<\over\sim}
\def\compoundrel#1\over#2{\mathpalette\compoundreL{{#1}\over{#2}}}
\def\compoundreL#1#2{\compoundREL#1#2}
\def\compoundREL#1#2\over#3{\mathrel
  {\vcenter{\hbox{$\m@th\buildrel{#1#2}\over{#1#3}$}}}}
\makeatother

\newcommand{\mapright}[1]{%
        \smash{\mathop{%
        \hbox to 1cm{\rightarrowfill}}\limits^{#1}}}

\newcommand{\ca}{\cos \alpha_{12}}

\begin{document}
\draft
\title{{SO(10) GUT and Quark-Lepton Mass Matrices}}

\author{K. MATSUDA and T. FUKUYAMA}
\address{
Department of Physics, 
Ritsumeikan University, Kusatsu, Shiga, 525-8577 Japan}
\author{H. NISHIURA}
\address{
Department of General Education, 
Junior College of Osaka Institute of Technology, \\
Asahi-ku, Osaka,535-8585 Japan}

\date{June 13th, 1999}
\maketitle

\begin{abstract}

The phenomenological model that all quark and lepton mass matrices have 
the 
same zero texture, namely their (1,1), (1,3) and (3,1) components are 
zeros, is discussed in the context of SO(10) Grand Unified Theories 
(GUTs). The mass matrices of type I for quarks are consistent with 
the experimental data in the quark sector. For the lepton sector, 
consistent fitting to the data of neutrino oscillation experiments 
force us to use the mass matrix for 
the charged leptons which is slightly deviated from type I.  
Given quark masses and charged lepton masses, the model includes 
19 free parameters,  
whereas the SO(10) GUTs gives 16 constrained equations.  Changing the 
remaining three parameters freely, 
we can fit all the entries of the CKM quark 
mixing matrix and the MNS lepton mixing matrix, and three neutrino 
masses consistently with the present experimental data.
\end{abstract}

\narrowtext
\section{Introduction}
The Downward and upward discrepancy in the atmospheric neutrino deficit in Super 
Kamiokande \cite{skamioka} together with other neutrino oscillation 
experiments such as solar neutrino \cite{skamioka2}, reactor\cite{chooz} 
and 
accelerator \cite{chorus} experiments drives us to the definite 
conclusion that neutrinos have masses.  These experiments enable us to 
get a glimpse of high energy physics beyond the Standard Model. 
In these situations our strategy is as follows.  First, we search for 
the most suitable phenomenological quark and lepton mass matrices which 
satisfies miscellaneous experiments in the hadron and electro-weak 
interactions.  Next, in order to search for its uniqueness and for its 
physical implications, such mass matrices are incorporated into the Grand 
Unified Theories (GUTs).
Of course phenomenological mass matrices and GUTs are closely correlated 
and the real model building is performed going back and forth between 
these two approaches.  Indeed, we 
consider the seesaw mechanism in neutrino mass matrix \cite{yanagida}, which 
supports minimally SO(10) GUTs. 
Conversely SO(10) GUTs prefer the mass matrices reflecting some 
similarity in the 
quark and lepton sectors.
In the seminal work of phenomenological quark mass matrix models 
\cite{fritzsch}, Fritzsch proposed a symmetric 
or hermitian matrices later called a six texture zero 
model which  has vanishing (1,1), (1,3), (3,1) and (2,2) components in both 
the mass matrices, $M_{u}$ for up-type quarks  (\(u,c,t\)) and $M_{d}$ 
for down-type quarks (\(d,s,b\)).  
Here n texture zero means that two types of quark mass 
matrices have totally n zeros in the upper half of hermitian mass 
matrices, in this case (1,1), (1,3) and (2,2) in each mass matrix.  
However, this model failed to predict a large top quark mass. Symmetric 
or hermitian six and five texture zero models were systematically 
discussed by Ramond et.al. \cite{ramond}.  They found that the hermitian 
$M_{u}$ and $M_{d}$  compatible with experiments can have at most five 
texture zero. Before the work of Ramond et.al. nonsymmetric or 
nonhermitian six texture zero quark  mass matrices model 
(nearest-neighber interaction (NNI) model) 
was proposed by  Branco-Lavoura-Mota \cite{branco}, and Takasugi showed 
that, by rebasing and rephasing of weak bases, always one of $M_{u}$ and 
$M_{d}$ can have the symmetric Fritzsch form and the other does NNI form 
\cite{takasugi}.
Demanding to deal with the quark and lepton mass matrices on the same footing, 
we have proposed 
a four texture zero model \cite{nishiura3}, in which all the quark and 
lepton mass matrices, $M_{u}, M_{d}, M_{e}$ and $M_{\nu}$ are hermitian 
and have the same textures. Here \(M_\nu\) and  \(M_e\) are mass 
matrices of neutrinos (\(\nu_e,\nu_\mu,\nu_\tau\)) and 
charged leptons (\(e,\mu,\tau\)), respectively.
Namely their (1,1), (1,3) and (3,1) components are zeros and the others 
are nonzero valued. This model was also discussed by Du and Xing \cite{du}, 
by Fritzsch and Xing \cite{fritzsch2}, by Kang and Kang \cite{kang}, 
by Kang, Kang, Kim, and Kim \cite{kang2}, and by 
Chkareuli and Froggatt \cite{chkareuli}, mainly in the quark sector.  
This model is compatible with the large top quark mass, the small quark mixing 
angles, and the large \(\nu_\mu\)-\(\nu_\tau\) neutrino mixing angles via the seesaw mechanism.  In this 
article, we discuss the above four texture zero model embedding in the SO(10) 
GUTs. The SO(10) GUTs impose some further constraints on the mass matrices. 
Using those constraints we predict all the entries of 
the lepton mixing matrix and neutrino masses, 
which are consistent with the experimental data, in terms of 
three free parameters left in the model.
\par
This article is organized as follows. In section 2 we review four texture 
zero model.  In section 3 we present a mass matrix model 
motivated by SO(10) GUTs . This model is combined with the four texture 
zero ansatzae in section 4. Section 5 is devoted to summary.

\par
\section{Four Texture Zero Quark-Lepton Mass Matrices}
Phenomenological quark mass matrices have been discussed from various 
points of view \cite{shizuoka}.  In this section we review our quark and lepton 
mass matrix model \cite{nishiura3}.  
The mass term in the Lagrangian is given by 
\begin{eqnarray}
L_M = 
-\overline{q_{R,i}^{u}}M_{uij}q_{L,j}^{u}
-\overline{q_{R,i}^{d}}M_{dij}q_{L,j}^{d}
-\overline{l_{R,i}}M_{eij}l_{L,j} 
-\overline{\nu'_{R,i}}M_{Dij}\nu_{L,j} \nonumber \\
-\frac{1}{2}\overline{(\nu_{L,i})^{c}}M_{Lij}\nu_{L,j}
-\frac{1}{2}\overline{(\nu'_{R,i})^{c}}M_{Rij}\nu'_{R,j}+H.c.
\end{eqnarray}
with
\begin{equation}
q_{L,R}^{u}=
\left(
	\begin{array}{c}
	u \\
	c \\
	t
	\end{array}
\right)_{L,R}, \quad
q_{L,R}^{d}=
\left(
	\begin{array}{c}
	d \\
	s \\
	b
	\end{array}
\right)_{L,R}, \quad
l_{L,R}=
\left(
	\begin{array}{c}
	e \\
	\mu \\
	\tau
	\end{array}
\right)_{L,R}, \quad
\nu_{L}=
\left(
	\begin{array}{c}
	\nu_e \\
	\nu_{\mu} \\
	\nu_{\tau}
	\end{array}
\right)_{L}, \quad
\nu'_{R}=
\left(
	\begin{array}{c}
	\nu'_e \\
	\nu'_{\mu} \\
	\nu'_{\tau}
	\end{array}
\right)_{R},
\end{equation}
where 
$M_u$, $M_d$, $M_e$, $M_D$, $M_L$, and $M_R$ are the mass matrices for up quarks, 
down quarks, charged leptons, Dirac neutrinos, 
left-handed Majorana neutrinos, and right-handed Majorana neutrinos, respectively.
The mass matrix of light Majorana neutrinos $M_\nu$ is given by 
\begin{equation}
M_\nu=M_L-M_D^T M_R^{-1} M_D, \label{eq051102}
\end{equation}
which is constructed via the seesaw mechanism 
\cite{yanagida} 
from the block-diagonalization of neutrino mass matrix,
\begin{equation}
\left(
        \begin{array}{cc}
        M_L & M_D^T \\
        M_D & M_R 
        \end{array}
\right).
\end{equation}
\par
We put a ansatz that the mass matrices $M_{u}, M_{d}, M_{e}$ and $M_{\nu}$ 
are hermitian and have 
the same textures.  Our model is  different from the Fritzsch model in the 
sense that (2,2) components are not zeros and that our model deals with 
the quark and lepton mass matrices on the same footing.  
The mass matrices \(M_D\), \(M_L\), and \(M_R\) are, furthermore, assumed 
to have the same zero 
texture as \(M_\nu\).  This ansatz restricts the texture 
forms \cite{nishiura3} and we 
choose 
the following our texture because it is most closely related with the NNI 
form \cite{branco}.
\begin{equation}
\mbox{NNI}:
\left(
        \begin{array}{ccc}
         0  & * &  0 \\
         {*}  & 0 &  *\\
         0  & * &  *
        \end{array}
\right), \hspace{3cm}
\mbox{{\bf Our Texture}}:
\left(
        \begin{array}{ccc}
         0  & * &  0 \\
         {*}  & * &  *\\
         0  & * &  *
        \end{array}
\right).
\end{equation}
The nonvanishing (2,2) component distinguishes our form from NNI's.
Thus the quark and lepton mass matrices are described as follows.
\begin{eqnarray}
&&M_{u}=
\left(
        \begin{array}{ccc}
         0    & A_{u} & 0 \\
        A_{u} & B_{u} & C_{u}\\
         0    & C_{u} & D_{u}
        \end{array}
\right), 
\nonumber \\
&&M_{d}=
P_d \left(
        \begin{array}{ccc}
         0  & A_{d} &  0 \\
        A_{d} & B_{d} & C_{d}\\
         0  & C_{d} & D_{d}
        \end{array}
\right) P_d^\dagger 
=
\left(
        \begin{array}{ccc}
         0  & A_{d} e^{i\alpha_{12}} &  0 \\
        A_{d} e^{-i\alpha_{12}} & B_{d} & C_{d} e^{i\alpha_{23}}\\
         0  & C_{d} e^{-i\alpha_{23}} & D_{d}
        \end{array}
\right),\nonumber\\
&&M_{e}=
P_e \left(
        \begin{array}{ccc}
         0  & A_{e} &  0 \\
        A_{e} & B_{e} & C_{e}\\
         0  & C_{e} & D_{e}
        \end{array}
\right) P_e^\dagger 
=
\left(
        \begin{array}{ccc}
         0  & A_{e} e^{i\beta_{12}} &  0 \\
        A_{e} e^{-i\beta_{12}} & B_{e} & C_{e} e^{i\beta_{23}}\\
         0  & C_{e} e^{-i\beta_{23}} & D_{e}
        \end{array}
\right),\label{eq051101}\\
&&M_{\nu}=
\left(
        \begin{array}{ccc}
         0  & A_{\nu} &  0 \\
        A_{\nu} & B_{\nu} & C_{\nu}\\
         0  & C_{\nu} & D_{\nu}
        \end{array}
\right), \nonumber 
\end{eqnarray}
where 
\(P_d\equiv\mbox{diag}(e^{i\alpha_1},e^{i\alpha_2},e^{i\alpha_3})\), 
\(\alpha_{ij}\equiv \alpha_i-\alpha_j\), and 
 \(P_e\equiv\mbox{diag}(e^{i\beta_1},e^{i\beta_2},e^{i\beta_3})\), 
\(\beta_{ij}\equiv \beta_i-\beta_j\).
\par
Let us discuss the relations between  the following texture's components 
of mass matrix \(M\):
\begin{equation}
{\normalsize M=}\left(
        \begin{array}{ccc}
        0 & A & 0\\
        A & B & C\\
        0 & C & D
        \end{array}
\right)
\end{equation}
and its eigenmass $m_i$. They satisfy
\begin{eqnarray}
B+D&=& m_1+m_2+m_3, \nonumber \\
BD-C^2-A^2&=&m_1m_2+m_2m_3+m_3m_1, \nonumber \\
DA^2&=&-m_1m_2m_3.\label{eq99020301}
\end{eqnarray}
Therefore, the mass matrix is classified into two types by 
choosing \(B\) and \(D\) as follows:
\begin{eqnarray}
&&\mbox{[type I]} \quad
\ B=m_2,\ D=m_3+m_1\nonumber\\
&&\mbox{[type II]} \quad
\ B=m_1, \ D=m_3+m_2
\end{eqnarray}
In the previous paper \cite{nishiura3} we showed that type I is 
compatible with the 
experimental data both for the quark and lepton mass matrices. 
So, we concentrate ourselves on the type I case. 
\par
In the type I case (\(B=m_2\), \(D=m_3+m_1\)), the other $A$ and $C$ 
take
the following value from Eq.(\ref{eq99020301})
\begin{equation}
A=\sqrt{\frac{(-m_1)m_2m_3}{m_3+m_1}}, \qquad
C=\sqrt{\frac{(-m_1)m_3(m_3-m_2+m_1)}{m_3+m_1}}.
\end{equation}
Transforming \(m_1\) into \(-m_1\)  by rephasing, 
the mass matrix \(M\) becomes
\begin{eqnarray}
&&M=\left(
        \begin{array}{ccc}
        0&\scriptstyle{\sqrt{\frac{m_1m_2m_3}{m_3-m_1}}}&0\\
        \scriptstyle{\sqrt{\frac{m_1m_2m_3}{m_3-m_1}}}
        & m_2 & 
    \scriptstyle{\sqrt{\frac{m_1m_3(m_3-m_2-m_1)}{m_3-m_1}}}\\
        0& \scriptstyle{\sqrt{\frac{m_1m_3(m_3-m_2-m_1)}{m_3-m_1}}}& 
    m_3-m_1
        \end{array}
\right) \simeq 
\left(
        \begin{array}{ccc}
        0& \sqrt{m_1m_2}& 0\\
        \sqrt{m_1m_2} & m_2 & \sqrt{m_1m_3}\\
        0& \sqrt{m_1m_3} & m_3-m_1
        \end{array}
\right). \nonumber \\
&&\hspace{10cm} (\mbox{for }m_3 \gg m_2 \gg m_1). \label{eq011601}
\end{eqnarray}
The orthogonal matrix \(O\) which diagonalize \(M\) in Eq.(\ref{eq011601}) as
\begin{equation}
O^T
\left(
        \begin{array}{ccc}
        0& \sqrt{m_1m_2}& 0\\
        \sqrt{m_1m_2} & m_2 & \sqrt{m_1m_3}\\
        0& \sqrt{m_1m_3} & m_3-m_1
        \end{array}
\right)
O=
\left(
        \begin{array}{ccc}
        -m_1 & 0 & 0\\
        0 & m_2 & 0\\
        0 &  0 & m_3
        \end{array}
\right),
\end{equation}
is given by 
\begin{eqnarray}
O& = &
\left(
\begin{array}{ccc}
{\sqrt{\frac{m_2m_3^2}{(m_2+m_1)(m_3^2-m_1^2)}}}&
{\sqrt{\frac{m_1m_3(m_3-m_2-m_1)}{(m_2+m_1)(m_3-m_2)(m_3-m_1)}}}&
{\sqrt{\frac{m_1^2m_2}{(m_3-m_2)(m_3^2-m_1^2)}}} \\
-{\sqrt{\frac{m_1m_3}{(m_2+m_1)(m_3+m_1)}}}&
{\sqrt{\frac{m_2(m_3-m_2-m_1)}{(m_2+m_1)(m_3-m_2)}}}&
{\sqrt{\frac{m_1m_3}{(m_3-m_2)(m_3+m_1)}}} \\
{\sqrt{\frac{{{m_1}^2}(m_3-m_2-m_1)}{(m_2+m_1)(m_3^2-m_2^2)}}}&
-{\sqrt{\frac{m_1m_2m_3}{(m_3-m_2)(m_2+m_1)(m_3-m_1)}}}&
{\sqrt{\frac{(m_3)^2(m_3-m_2-m_1)}{(m_3^2-m_2^2)(m_3-m_2)}}} 
\end{array}
\right) \nonumber \\
&\simeq&
\left(
        \begin{array}{ccc}
        1& \displaystyle{\sqrt{\frac{m_1}{m_2}}}&
         \scriptstyle{\sqrt{\frac{m_1m_2^2}{m_3^3}}}\\
        \displaystyle{-\sqrt{\frac{m_1}{m_2}}}
        & 1 & \displaystyle{\sqrt{\frac{m_1}{m_3}}}\\
        \displaystyle{\sqrt{\frac{{\scriptstyle{m_1^2}}}{m_2m_3}}}
        & \displaystyle{-\sqrt{\frac{m_1}{m_3}}} & 1
        \end{array}
\right) \qquad (\mbox{for }m_3 \gg m_2 \gg m_1).
\label{eq990114}
\end{eqnarray}
The mass matrices for quarks and charged leptons, 
\(M_d\), \(M_u\), and \(M_e\) are
considered to be of this type I and are given by 
\begin{eqnarray}
M_d &\simeq&
P_d
\left(
        \begin{array}{ccc}
        0              &  \sqrt{m_d m_s}  &  0\\
        \sqrt{m_d m_s} &  m_s               &  \sqrt{m_d m_b}\\
        0              &  \sqrt{m_d m_b}  &  m_b-m_d
        \end{array}
\right)P_d^\dagger, \quad
M_u \simeq
\left(
        \begin{array}{ccc}
        0              &  \sqrt{m_u m_c}  &  0\\
        \sqrt{m_u m_c} &  m_c               &  \sqrt{m_u m_t}\\
        0              &  \sqrt{m_u m_t}  &  m_t-m_u
        \end{array}
\right), \nonumber\\
M_e &\simeq&
P_e
\left(
        \begin{array}{ccc}
        0              &  \sqrt{m_e m_\mu}  &  0\\
        \sqrt{m_e m_\mu} &  m_\mu              &  \sqrt{m_e m_\tau}\\
        0              &  \sqrt{m_e m_\tau}  &  m_\tau-m_e
        \end{array}
\right)P_e^\dagger. 
\end{eqnarray}
Those \(M_d\), \(M_u\), and \(M_e\) are, respectively, diagonalized by 
matrices \(P_d O_d\), \(O_u\), and \(P_e O_e\). 
Here the orthogonal matrices \(O_d\), \(O_u\) and \(O_e\) which diagonalize
\(P_d^\dagger M_d P_d\), \(M_u\), and \(P_e^\dagger M_e P_e\) 
are obtained from Eq. (\ref{eq990114}) 
by replacing \(m_1\), \(m_2\), \(m_3\) by \(m_d\), \(m_s\), \(m_b\) 
, by \(m_u\), \(m_c\), \(m_t\), and by \(m_e\), \(m_\mu\), \(m_\tau\), 
respectively.
In this case, the Cabibbo-Kobayashi-Maskawa (CKM) quark mixing matrix 
\(V\) can be written as
\begin{equation}
V=P_q^{-1} P_d^{-1} O_u^T P_d O_d P_q 
\simeq
\left(
        \begin{array}{ccc}
         |V_{11}| & |V_{12}| & |V_{13}| e^{-i \phi}\\
        -|V_{12}| & |V_{22}| & |V_{23}|\\
         |V_{12}V_{23}|-|V_{13}|e^{i \phi}\ & -|V_{23}| & |V_{33}|\\
        \end{array}
\right). \label{eq99012501}
\end{equation}
where the \(P_d^{-1}\) factor is included to put \(V\) in the form
with diagonal elements real to a good approximation.
Furthermore, the \(P_q^{-1}\) and 
\(P_q=\mbox{diag}(e^{i\phi_1},e^{i\phi_2},e^{i\phi_3})\) with 
\(\phi_1-\phi_2=\mbox{arg}(P_d^{-1} O_u^T P_d O_d)_{12}\) and 
\(\phi_1-\phi_3=\mbox{arg}(P_d^{-1} O_u^T P_d O_d)_{23}\) 
come from the choice of phase convention as Eq. (\ref{eq99012501}).
The explicit forms  of the 
components of \(V\) are obtained \cite{nishiura3} as
\begin{eqnarray}
&&|V_{12}|\simeq
\left|
\sqrt{\frac{m_d}{m_s}}-\sqrt{\frac{m_u}{m_c}}e^{-i\alpha_{12}}
\right|,
\nonumber \\
&&|V_{23}|\simeq
\left|
\sqrt{\frac{m_d}{m_b}}-\sqrt{\frac{m_u}{m_t}}e^{-i\alpha_{23}}
\right|,
\nonumber \\
&&|V_{13}|\simeq
\left|
\sqrt{\frac{m_d^2 m_s}{m_b^3}}-
\sqrt{\frac{m_u}{m_c}}
\left(\sqrt{\frac{m_d}{m_b}}-\sqrt{\frac{m_u}{m_t}}e^{-i\alpha_{23}}
\right)e^{-i\alpha_{12}}
\right|\nonumber \\
&&\cos\phi \simeq 
\frac{|V_{12}|^2+m_u/m_c-m_d/m_s}{2|V_{12}|\sqrt{m_u/m_c}}.
\label{eq122003}
\end{eqnarray}

The lepton mixing matrix \(U\) 
[hereafter we call it 
the Maki-Nakagawa-Sakata (MNS) mixing matrix \cite{MNS}], 
is given by
\begin{equation}
U=P_e^\dagger O_e^T P_e O_\nu=
 \left(
 	\begin{array}{ccc}
 	U_{11} & U_{12} & U_{13}\\
 	U_{21} & U_{22} & U_{23}\\
 	U_{31} & U_{32} & U_{33}
 	\end{array}
 \right), \label{123002}
\end{equation}
where the \(P_e^\dagger\) factor is included to put \(U\) in the form with diagonal
elements real to a good approximation. 
Here the \(O_\nu\) is the orthogonal matrix which diagonalizes 
the light Majorana neutrino mass matrices 
\(M_\nu\) given by Eq.(\ref{eq051102}).

\par
\section{Mass Matrices in the context of SO(10) GUTs}
Even if we succeeded in constructing the quark mass matrices $M_u$ and $M_d$ 
consistent with experiments, we have infinitely many mass matrices equivalent 
to the $M_u$ and $M_d$ which are defined as
\begin{equation}
M_u'=F^\dagger M_uG_u \quad M_d'=F^\dagger M_dG_d,
\end{equation}
with arbitrary unitary matrices $F$, $G_u$, and $G_d$ in the standard 
$SU_L(2)\times U_Y(1)$ model, and with $G_u = G_d$ in the 
$SU_L(2)\times SU_R(2)\times U_Y(1)$ model.
The fact that quark and lepton mass matrices have the same form strongly 
suggests that the quarks and leptons belong to the same multiplets.  So in 
this section we try to incorporate our mass matrix in the context of 
SO(10) GUTs. 
We consider two SO(10) symmetry breaking patterns.
\begin{eqnarray}
\mbox{(i)}\ SO(10) & \rightarrow& SU(4)\times SU_L(2)\times SU_R(2) 
\rightarrow SU_c(3)\times SU_L(2)\times SU_R(2)\times U(1) 
\rightarrow G_s, \nonumber \\
\mbox{(ii)}\ SO(10) & \rightarrow& SU(5) 
\rightarrow G_s ,  \label{eq01011}
\end{eqnarray}
where $G_s= SU_c(3)\times SU_L(2)\times U(1)$.

\subsection{The case of $SO(10$) breaking down to $SU(4)\times SU_L(2)\times SU_R(2)$}
Here we consider the charge-conjugation-conserving (CCC) 
version \cite{bottino}\cite{johnson}\cite{bottino2} of the SO(10) model 
in which Left-Right discrete (not manifest) symmetry is imposed. 
\par 
In the SO(10) model\cite{chanowitz} \cite{Harvey} \cite{Vergados}, the left- (right-) handed fermions 
$\psi_{L(R)i}$ in a given i-th generation are assigned to a single irreducible 
{\bf 16}.
Since ${\bf 16}\times {\bf 16} = {\bf 10}_S +{\bf 120}_A + 
{\bf 126}_S$,  the fermion  masses are 
generated when the Higgs fields of {\bf 10}, and {\bf 120}, and 
{\bf 126} dimensional SO(10) representation (denoted by 
$\phi_{10}$, $\phi_{120}$, and $\phi_{126}$, respectively) develop 
nonvanishing expectation values. Their decomposition under 
$SU(4)\times SU_L(2)\times SU_R(2)$ are given by 
\begin{eqnarray}
{\bf 10}  &=& ({\bf 6},{\bf 1},{\bf 1}) +
	({\bf 1},{\bf 2},{\bf 2}), \nonumber \\
{\bf 120} &=& ({\bf 15},{\bf 2},{\bf 2})+({\bf 6},{\bf 3},{\bf 1})+
	(\overline{{\bf 6}},{\bf 1},{\bf 3})+({\bf 1},{\bf 2},{\bf 2})+
		({\bf 20},{\bf 1},{\bf 1}), \\
{\bf 126} &=& ({\bf 10},{\bf 3},{\bf 1})+
	(\overline{{\bf 10}},{\bf 1},{\bf 3}) +
		({\bf 15},{\bf 2},{\bf 2})+
		(\overline{{\bf 6}},{\bf 1},{\bf 1}). \nonumber
\end{eqnarray}
On the other hand, the fermion field of 16-dimensinal SO(10) representation is 
decomposed as 
\begin{eqnarray}
{\bf 16} & =({\bf 4},{\bf 2},{\bf 1})+(\overline{{\bf 4}},{\bf 1},{\bf 2}).
\end{eqnarray}
With respect to $SU(4)\times SU_L(2)\times SU_R(2)$, the left - and right- handed  
quarks and leptons of a given i-th generation are assigned as
\begin{eqnarray}
	\left(
		\begin{array}{cccc}
		u_r & u_y & u_b & \nu_e \\
		d_r & d_y & d_b & e
		\end{array}
	\right)_{L(R)}	\equiv F_{L(R)1} ,
\end{eqnarray}
\(F_{L(R)2}\) and \(F_{L(R)3}\) are likewise defined for 
the 2nd and 3rd generations.
Note that their transformation properties are
$F_{Li}=({\bf 4},{\bf 2},{\bf 1})$ 
and $F_{Ri}=({\bf 4},{\bf 1},{\bf 2})$ 
and that ($F_{Li} + \overline{F_{Ri}}$ ) 
yields the {\bf 16} of SO(10).
Since $({\bf 4},{\bf 2},{\bf 1})\times
(\overline{{\bf 4}},{\bf 1},{\bf 2})=
({\bf 15},{\bf 2},{\bf 2})+({\bf 1},{\bf 2},{\bf 2})$, 
the Dirac masses for quarks and leptons are generated when neutral components 
in a ({\bf 1},{\bf 2},{\bf 2}) multiplet in $\phi_{10}$, 
({\bf 1},{\bf 2},{\bf 2}) and ({\bf 15},{\bf 2},{\bf 2}) in 
$\phi_{120}$, and ({\bf 15},{\bf 2},{\bf 2}) in $\phi_{126}$ of 
$SU(4)\times SU_L(2)\times SU_R(2)\subset SO(10)$ 
develop nonvanishing expectation values.  
On the other hand, the $(\overline{{\bf 10}},{\bf 3},{\bf 1})$ and 
$({\bf 10},{\bf 1},{\bf 3})$ in $\phi_{126}$ are responsible 
for the left- and the right- handed Majorana neutrino masses 
through the Higgs-lepton-lepton interactions  
$(\overline{{\bf 10}},{\bf 3},{\bf 1})({\bf 4},{\bf 2},{\bf 1})
({\bf 4},{\bf 2},{\bf 1})$ and 
$({\bf 10},{\bf 1},{\bf 3})
(\overline{{\bf 4}},{\bf 1},{\bf 2})
(\overline{{\bf 4}},{\bf 1},{\bf 2})$, respectively. 
Here the $(\overline{{\bf 10}},{\bf 3},{\bf 1})$ is the $SU_L(2)$ 
triplet Higgs (denoted by $\phi(\overline{{\bf 10}},{\bf 3},{\bf 1})$) 
and the $(\overline{{\bf 10}},{\bf 3},{\bf 1})$ is the $SU_R(2)$ 
triplet Higgs ($\phi({\bf 10},{\bf 1},{\bf 3})$).

In the CCC version of the SO(10) model , 
the mass matrices 
$M_u$, $M_d$, $M_e$, $M_D$, $M_L$, and $M_R$, for up quarks, 
down quarks, charged leptons, Dirac neutrinos, 
left-handed Majorana neutrinos, and right-handed Majorana neutrinos, respectively, 
are given, in the lowest tree level, by  
\begin{eqnarray}
M_u & =& S^{(10)}\langle\phi^1_+\rangle
  + A^{(120)}(\langle\phi^3_+\rangle+\frac{1}{3}\langle\phi^{3 
\prime}_+\rangle)
  +S^{(126)}\frac{1}{3}\langle\phi^{5}_+\rangle, \nonumber \\
M_d & =& S^{(10)}\langle\phi^1_-\rangle
  + A^{(120)}(-\langle\phi^3_-\rangle+\frac{1}{3}\langle\phi^{3 
\prime}_-\rangle)
  -S^{(126)}\frac{1}{3}\langle\phi^{5}_-\rangle, \nonumber \\
r M_e & =& S^{(10)}\langle\phi^1_-\rangle
  + A^{(120)}(-\langle\phi^3_-\rangle-\langle\phi^{3 \prime}_-\rangle)
  +S^{(126)}\langle\phi^{5}_-\rangle, \label{eq05010} \\
r' M_D & =& S^{(10)}\langle\phi^1_+\rangle
  + A^{(120)}(\langle\phi^3_+\rangle-\langle\phi^{3 \prime}_+\rangle)
  -S^{(126)}\langle\phi^{5}_+\rangle, \nonumber \\
s M_L & =& S^{(126)}\langle\phi(\overline{{\bf 10}},{\bf 3},{\bf 1})\rangle, \nonumber \\
s' M_R & =& S^{(126)}\langle\phi({\bf 10},{\bf 1},{\bf 3})\rangle, \nonumber
\end{eqnarray}
where $\langle\phi^1_\pm\rangle$ are the vacuum expectation values of 
the Higgs fields of $\phi_{10}$,
$\langle\phi^3_\pm\rangle$ and $\langle\phi^{3 \prime}_\pm\rangle$ of 
$\phi_{120}$, and $\langle\phi^5_\pm\rangle$, 
$\langle\phi(\overline{{\bf 10}},{\bf 3},{\bf 1})\rangle$ and 
$\langle\phi({\bf 10},{\bf 1},{\bf 3})\rangle$ of $\phi_{126}$. 
See Ref\cite{bottino2} for details and the notations.
The matrices $S^{(10)}$ and $S^{(126)}$ are real symmetric matrices and 
$A^{(120)}$is a pure imaginary matrix. These matrices are 
the coupling-constant matrices which appear in the Yukawa coupling of 
fermion fields with Higgs field , which is given by 
\begin{eqnarray}
2L_Y & =& S_{ij}^{(10)}\overline{(\psi_{Li})^c}\phi_{10}\psi_{Lj}
	 + A_{ij}^{(120)}\overline{(\psi_{Li})^c}\phi_{120}\psi_{Lj} 
	+ S_{ij}^{(126)}\overline{(\psi_{Li})^c}\phi_{126}\psi_{Lj} 
	+ (L \leftrightarrow R) +H.c. \label{eq05011}
\end{eqnarray}
The $\psi_{L(R)i}$ are the 16 irreducible 
representations of the left-  and right- handed fermion fields in a 
given i'th generation.
The property that  $S^{(10)}$ and $S^{(126)}$ are symmetric and 
$A^{(120)}$ is antisymmetric 
results from the decomposition 
${\bf 16}\times {\bf 16} = {\bf 10}_S +{\bf 120}_A + {\bf 126}_S$, 
whereas the property that  $S^{(10)}$ and $S^{(126)}$ are real 
and $A^{(120)}$ is pure imaginary is a 
consequence of their 
being Hermitian, which in turn comes from the requirement of the 
invariance of $L_Y$ under 
the discrete symmetry $\psi_L \leftrightarrow \psi_R^c$ \cite{bottino2}. 
In Eq.(\ref{eq05010}), the factors $r\simeq(2\sim 3)$, $r'$, $s$ and $s'$ 
, all roughly of order unity, 
are the renormarization-group-equation factors \cite{buras} \cite{johnson} 
which arise from 
the differences in the renormalization of the lepton and quark masses 
due to the color 
quantum numbers of the quarks and so on.
The overall factor comes from the loop correction of gauge boson 
in the renormalization group equation.  
Exactly we should consider the evolution equation of Yukawa coupling and 
in this case mass matrices gets renormalized in somewhat different form.
Therefore, this form is an approximation.  In this point we will also 
discuss in the last section.
\par
We now make the following assumptions.
\par
(i) The contribution from {\bf 120} is assumed to be small compared with the 
contributions from 
{\bf 10} and {\bf 126}, and hence it is neglected in 
$M_u $ and $M_D$. On the other hand, it is retained 
in $M_d$ and $M_e$, for the main term $S^{(10)}\langle\phi^1_-\rangle$ 
is smaller by the 
factor $\alpha=\langle\phi_-^1\rangle/\langle\phi_+^1\rangle$, which is 
of order of $(m_b/m_t)$ [see Eq.(\ref{eq042205})].
This is an assumption for simplicity 
in order to incorporate Eq.(\ref{eq051101})
\par
(ii) All the vacuum expectation values of Higgs fields are assumed to be 
real so that 
all the fermion mass matrices are Hermitian.

With these assumptions, Eqs. (\ref{eq05011}) becomes
\begin{eqnarray}
M_u & =& S+ \epsilon S', \nonumber \\
M_d & =& \alpha S + S' + A'=\alpha M_u+A'-(\alpha\epsilon-1)S', \nonumber \\
r M_e &=& \alpha S -3 S' + \delta A'=\alpha M_u +\delta A'-(\alpha\epsilon+3)S',
  \label{eq042201} \\ 
r' M_D &=& S - 3 \epsilon S', \nonumber \\
s M_L &=& \beta S', \nonumber \\ 
s' M_R &=& \gamma S'. \nonumber 
\end{eqnarray}
where the matrices $S$, $S'$ and $A'$ and the real parameters $\alpha$, 
$\beta$, $\gamma$ and $\delta$
 are defined by 
\begin{eqnarray}
S &=& S^{(10)}\langle\phi^1_+\rangle, \hspace{3.2cm}
S' = S^{(126)}(-\frac{1}{3}\langle\phi^5_-\rangle), \nonumber \\
A' &=& A^{(120)}(-\langle\phi^3_-\rangle+\frac{1}{3}\langle\phi^{3 
\prime}_-\rangle), \label{eq042202} \\
\alpha &=& \langle\phi_-^1\rangle/\langle\phi_+^1\rangle, \hspace{3.1cm} 
\beta =
\langle\phi(\overline{{\bf 10}},{\bf 3},{\bf 1})\rangle/(-\frac{1}{3}\langle\phi^{5}_-\rangle ), \nonumber \\ 
\gamma &=& \langle\phi({\bf 10},{\bf 1},{\bf 3})\rangle 
/(-\frac{1}{3}\langle\phi^{5}_-\rangle ), \qquad
\delta = (\langle\phi^3_-\rangle + \langle\phi^{3 \prime}_-\rangle)/ 
(\langle\phi^3_-\rangle-\frac{1}{3}\langle\phi^{3 \prime}_-\rangle). 
\nonumber 
\end{eqnarray}
Note that solving diagonal elements of Eq.(\ref{eq042201}) for $\alpha$, 
one finds 
\begin{equation}
 \alpha = \frac{3\mbox{Tr}M_d+r\mbox{Tr}M_e}{3\mbox{Tr}M_u+r'\mbox{Tr}M_D} 
 \simeq \frac{m_b}{m_t} \label{eq042205},
\end{equation}
which is about 0.02. As mentioned already, this is why the $ A^{(120)} $ 
and $S^{(126)}$ terms are kept 
in $M_d$ and $M_e$. 
The Eq.(\ref{eq042201}) is our SO(10)-motivated model 
for fermion mass matrices.

\subsection{The case of $SO(10$) breaking down to $SU(5)$}
In this case, the fermion  masses are also generated 
when the Higgs fields of {\bf 10}, and {\bf 120}, and {\bf 126} dimensional 
SO(10) representation (denoted by 
$\phi_{10}$, $\phi_{120}$, and $\phi_{126}$, respectively) develop 
nonvanishing expectation values. Their decomposition under 
$SU(5)$ are given by 
\begin{eqnarray}
{\bf 10}  &=& {\bf 5}+\overline{{\bf 5}},\nonumber \\
{\bf 120} &=& {\bf 5}+\overline{{\bf 5}}+{\bf 10}+
	\overline{{\bf 10}}+{\bf 45}+\overline{{\bf 45}},\\
{\bf 126} &=& {\bf 1}+\overline{{\bf 5}}+{\bf 10}
	+\overline{{\bf 15}}+{\bf 45}+{\bf 50}.\nonumber
\end{eqnarray}
The yukawa couplings in $L_Y$ gives the following fermion masses 
when the neutral components 
in a ${\bf 5}$ and $\overline{{\bf 5}}$ Higgs multiplets in 
$\phi_{10}$, ${\bf 5}$, $\overline{{\bf 5}}$, ${\bf 45}$, 
and $\overline{{\bf 45}}$ in $\phi_{120}$, and ${\bf 1}$, 
$\overline{{\bf 5}}$, $\overline{{\bf 15}}$, and ${\bf 45}$ 
in $\phi_{126}$ of $SU(5)\subset SO(10)$ 
develop nonvanishing expectation values. \cite{barbieri} \cite{Vergados} 
\begin{eqnarray}
S_{ij}^{(10)}\overline{(\psi_{Li})^c}\phi_{10}\psi_{Lj}  &\rightarrow&
	S_{ij}^{(10)}\{\langle\phi_{10}({\bf 5})
	\rangle(\overline{u_{R,i}}u_{L,j} + 
	\overline{\nu'_{R,i}}\nu_{L,j}) 
	+ \langle\phi_{10}(\overline{{\bf 5}})
	\rangle(\overline{d_{R,i}}d_{L,j} + \overline{e_{R,i}}e_{L,j})\},
	\nonumber \\
A_{ij}^{(120)}\overline{(\psi_{Li})^c}\phi_{120}\psi_{Lj}  &\rightarrow&
	A_{ij}^{(120)}\{ \langle\phi_{120}(\overline{{\bf 5}})\rangle
	(\overline{d_{R,i}}d_{L,j} + \overline{e_{R,i}}e_{L,j}) \nonumber \\
 	&&+\langle\phi_{120}(\overline{{\bf 45}})\rangle
 	(\overline{d_{R,i}}d_{L,j} -3 \overline{e_{R,i}}e_{L,j}) 
	+\langle\phi_{120}({\bf 5})\rangle
	\overline{\nu'_{R,i}}\nu_{L,j} \nonumber \\
	&&+\langle\phi_{120}({\bf 45})\rangle
	\overline{u_{R,i}}u_{L,j}  \},\nonumber \\
S_{ij}^{(126)}\overline{(\psi_{Li})^c}\phi_{126}\psi_{Lj},&\rightarrow& 
	S_{ij}^{(126)}\{\langle\phi_{126}({\bf 5})\rangle
	(\overline{u_{R,i}}u_{L,j} -3 \overline{\nu_{R,i}}\nu_{L,j}) \nonumber \\
    &&+\langle\phi_{126}(\overline{{\bf 45}})\rangle
    (\overline{d_{R,i}}d_{L,j} -3 \overline{e_{R,i}}e_{L,j}) 
    +\langle\phi_{126}({\bf 1})\rangle\overline{\nu'^c_{R,i}}\nu'_{R,j}
     \nonumber \\
	&&+\langle\phi_{126}({\bf 15})\rangle\overline{\nu^c_{L,i}}\nu_{L,j} \},
	\nonumber \\
\end{eqnarray}
Therefore, the mass matrices 
$M_u$, $M_d$, $M_e$, $M_D$, $M_L$, and $M_R$, for up quarks, 
down quarks, charged leptons, Dirac neutrinos, 
left-handed Majorana neutrinos, and right-handed Majorana neutrinos, respectively, 
are given by  
\begin{eqnarray}
M_u & =& S^{(10)}\langle\phi_{10}({\bf 5})\rangle
  + A^{(120)}\langle\phi_{120}({\bf 45})\rangle
  +S^{(126)}\langle\phi_{126}({\bf 5})\rangle, \nonumber \\
M_d & =& S^{(10)}\langle\phi_{10}(\overline{{\bf 5}})\rangle
  + A^{(120)}(\langle\phi_{120}(\overline{{\bf 5}})\rangle+\langle\phi_{120}(\overline{{\bf 45}})\rangle)
  +S^{(126)}\langle\phi_{126}(\overline{{\bf 45}})\rangle, \nonumber \\
r M_e & =& S^{(10)}\langle\phi_{10}(\overline{{\bf 5}})\rangle
  + A^{(120)}(\langle\phi_{120}(\overline{{\bf 5}})\rangle-3\langle\phi_{120}(\overline{{\bf 45}}))
  -3S^{(126)}\langle\phi_{126}(\overline{{\bf 45}})\rangle, \\
r' M_D & =& S^{(10)}\langle\phi_{10}({\bf 5})\rangle 
  + A^{(120)}\langle\phi_{120}({\bf 5})\rangle
  -3 S^{(126)}\langle\phi_{126}({\bf 5})\rangle, \nonumber \\
s M_L & =& S^{(126)}\langle\phi_{126}({\bf 15})\rangle, \nonumber \\
s' M_R & =& S^{(126)}\langle\phi_{126}({\bf 1})\rangle, \nonumber
\end{eqnarray}
These mass matrices reduce to the same form as Eq.(\ref{eq042201}) 
by assuming again that the contributions 
from {\bf 120} Higgs in $M_u$ and $M_D$ are negligible 
and by defining the matrices $S$, $S'$, and $A'$ and the real parameters $\alpha$, 
$\beta$, $\gamma$, and $\delta$ , instead of Eq.(\ref{eq042202}), as 
\begin{eqnarray}
S & =&  S^{(10)}\langle\phi_{10}({\bf 5})\rangle, \hspace{3.5cm}
S'  = S^{(126)}\langle\phi_{126}(\overline{{\bf 45}})\rangle, \nonumber \\
A' &=& A^{(120)}(\langle\phi_{120}(\overline{{\bf 5}})\rangle
     +\langle\phi_{120}(\overline{{\bf 45}})\rangle),  \nonumber \\
\alpha &=& \langle\phi_{10}(\overline{{\bf 5}})\rangle/
\langle\phi_{10}({\bf 5})\rangle, \hspace{2.8cm}
\beta = 
\langle\phi_{126}({\bf 15})\rangle/
   \langle\phi_{126}(\overline{{\bf 45}})\rangle,  \\ 
\gamma &=& \langle\phi_{126}({\bf 1})\rangle/
\langle\phi_{126}(\overline{{\bf 45}})\rangle,  \nonumber \\
\delta &=& (\langle\phi_{120}(\overline{{\bf 5}})\rangle-3
\langle\phi_{120}(\overline{{\bf 45}}))/ 
(\langle\phi_{120}(\overline{{\bf 5}})\rangle+
\langle\phi_{120}(\overline{{\bf 45}})\rangle),
\nonumber 
\end{eqnarray}
Thus, Eq.(\ref{eq042201}) is our SO(10)-motivated model 
for fermion mass matrices both for the two SO(10) breaking patterns (i) and (ii) in Eq.(\ref{eq01011}). 

\par
\section{Four texture zero model in SO(10)}
The SO(10) model Eq.(\ref{eq042201}) is now combined with the four 
texture zero ansatzae for 
$M_u$, $M_d$ and $M_e$ which are given by Eq.(\ref{eq051101}). 

\par
First it follows from Eq.(\ref{eq042201}) that $S$, $S'$ and 
$A'$ are represented in terms of the symmetric (antisymmetric) parts,
\(M_{sym}\) (\(M_{antisym}\)), of \(M_u\), \(M_d\) and \(M_e\);
\begin{eqnarray}
(1-\alpha\epsilon)S & =& (M_u)_{sym}-\epsilon(M_d)_{sym}, \nonumber \\
S' & =& \frac{1}{4}\{(M_d)_{sym}-r(M_e)_{sym}\}, \label{eq05200}  \\ 
A' &=& (M_d)_{antisym}. \nonumber 
\end{eqnarray}
We also find the constraints 
\begin{eqnarray}
(1-\alpha\epsilon)r(M_e)_{sym} & =& 4\alpha (M_u)_{sym}
	-(3+\alpha\epsilon)(M_d)_{sym}, \nonumber \\
\delta (M_d)_{antisym} &=& r(M_e)_{antisym}. \label{eq05333}  
\end{eqnarray}
Using the four texture zero ansatzae for 
$M_u$, $M_d$ and $M_e$  given by Eq.(\ref{eq051101}), 
the respective elements of Eq.(\ref{eq05333}) become
\begin{eqnarray}
(1-\alpha\epsilon)rA_e \cos\beta_{12} & =& 
4\alpha A_u-(3+\alpha\epsilon)A_d \cos\alpha_{12}, \nonumber \\
(1-\alpha\epsilon)rB_e & =& 4\alpha B_u-(3+\alpha\epsilon)B_d, \nonumber 
\\
(1-\alpha\epsilon)rC_e \cos\beta_{23} & =& 
4\alpha C_u-(3+\alpha\epsilon)C_d \cos\alpha_{23}, \nonumber \\
(1-\alpha\epsilon)rD_e & =& 
4\alpha D_u-(3+\alpha\epsilon)D_d,  \label{eq05444} \\
\delta A_d \sin\alpha_{12} &=& 
rA_e \sin\beta_{12}, \nonumber \\
\delta C_d \sin\alpha_{23} &=& 
rC_e \sin\beta_{23}. \nonumber
\end{eqnarray}
In Eq.(\ref{eq05444}) there are six equations and eight unknown parameters, 
namely $\alpha$, $\epsilon$, $\delta$, $\alpha_{12}$, 
$\alpha_{23}$, $\beta_{12}$, $\beta_{23}$ and $r$ 
provided that $A_u$, $B_u$,...,$D_e$ are given. 
In the following, we treat $\cos\alpha_{12}$ and $r$ as free 
parameters so that 
all the other parameters are functions of them. 
Here we treat $r$ as a free parameter too, although we know $r \simeq (2\sim 3)$. 
Let us present the following useful expressions which are derived from Eq(\ref{eq05444}):
\begin{eqnarray}
&&\cos\beta_{12} = \left(\frac{B_eD_d-D_eB_d}{B_uD_d-D_uB_d}\right)\left(\frac{A_u}{A_e}\right) 
	- \left(\frac{B_eD_u-D_eB_u}{B_uD_d-D_uB_d}\right)\left(\frac{A_d}{A_e}\right)\cos\alpha_{12}, \nonumber \\
&&\cos\beta_{23} = \left(\frac{B_eD_d-D_eB_d}{B_uD_d-D_uB_d}\right)\left(\frac{C_u}{C_e}\right) 
	- \left(\frac{B_eD_u-D_eB_u}{B_uD_d-D_uB_d}\right)\left(\frac{C_d}{C_e}\right)\cos\alpha_{23}, \label{eq04190}  \nonumber \\ 
&&\frac{\sin\beta_{23}}{\sin\alpha_{23}} 
 = \left(\frac{A_eC_d}{A_dC_e}\right)\frac{\sin\beta_{12}}{\sin\alpha_{12}},
   \nonumber \\
&&\alpha  =  \frac{r\left(\frac{B_eD_d-D_eB_d}{B_uD_d-D_uB_d}\right)}{r\left(\frac{B_eD_u-D_eB_u}{B_uD_d-D_uB_d}\right)+1}, 
\quad \epsilon = \frac{r\left(\frac{B_eD_u-D_eB_u}{B_uD_d-D_uB_d}\right)-3}{r\left(\frac{B_eD_d-D_eB_d}{B_uD_d-D_uB_d}\right)}, \quad 
\delta = r\left(\frac{A_e}{A_d}\right)\frac{\sin\beta_{12}}{\sin\alpha_{12}}. 
\label{eq090701}
\end{eqnarray}
\par 
Now we discuss the MNS lepton mixing matrix and neutrino masses. 
The light Majorana neutrino mass matrix $M_{\nu}$ is 
given by Eq.(\ref{eq051102}),
where the Dirac neutrino, left- and right- handed Majorana neutrino mass matrices 
$M_D$, $M_L$, and $M_R$ are expressed in terms of the entries 
of the quarks and charged lepton mass matrices due to the SO(10) 
constraints and their expressions are given, from Eqs.(\ref{eq042201}) and 
(\ref{eq05200}),  by 
\begin{eqnarray}
r' M_D & =& S - 3 \epsilon S' 
= \left(
	\begin{array}{ccc}
		0 & A_D & 0 \\ 
		A_D & B_D & C_D \\
		0 & C_D & D_D 
	\end{array}
\right), \nonumber \\ 
s M_L & =&  \beta S'
=\beta  \left(
		\begin{array}{ccc}
		0 & A_{S'} & 0 \\ 
		A_{S'} & B_{S'} & C_{S'} \\
		0 & C_{S'} & D_{S'} 
		\end{array}
	\right), \label{eq05400}  \\ 
s' M_R & =&  \gamma S' 
=\gamma  \left(
		\begin{array}{ccc}
		0 & A_{S'} & 0 \\ 
		A_{S'} & B_{S'} & C_{S'} \\
		0 & C_{S'} & D_{S'} 
		\end{array}
	\right), \nonumber 
\end{eqnarray}
where the elements $A_D$, $B_D$, $C_D$, $C_D$, $A_{S'}$, $B_{S'}$, 
$C_{S'}$, and $C_{S'}$  are obtained as
\begin{eqnarray}
A_D & =& A_u-\epsilon(A_d\cos\alpha_{12}-rA_e\cos\beta_{12}), \nonumber \\
B_D & =& B_u-\epsilon(B_d-rB_e), \nonumber \\
C_D & =& C_u-\epsilon(C_d\cos\alpha_{23}-rC_e\cos\beta_{23}), \label{eq05410}  \\
D_D & =& D_u-\epsilon(D_d-rD_e), \nonumber 
\end{eqnarray}
and 
\begin{eqnarray}
A_{S'} & =& \frac{1}{4}(A_d\cos\alpha_{12}-rA_e\cos\beta_{12}), \nonumber \\
B_{S'} & =& \frac{1}{4}(B_d-rB_e), \nonumber \\
C_{S'} & =& \frac{1}{4}(C_d\cos\alpha_{23}-rC_e\cos\beta_{23}), \label{eq05420}  \\
D_{S'} & =& \frac{1}{4}(D_d-rD_e). \nonumber 
\end{eqnarray}
In the following analysis, we assume that the contribution 
of $M_L$ to $M_{\nu}$ in Eq. (\ref{eq051102}) 
is much smaller than that of the second term 
so that we have $M_\nu=M_L-M_D^T M_R^{-1} M_D \simeq -M_D^T M_R^{-1} M_D$.
Then, all the components $A_{\nu}$, $B_{\nu}$, $C_{\nu}$, 
and $D_{\nu}$ in Eq.(\ref{eq051101}) 
are determined, from Eqs.(\ref{eq05400}), (\ref{eq05410}), and (\ref{eq05420}), 
as functions of 
$\cos\alpha_{12}$ and $r$ except for the common overall factor $s'/(r'^2 \gamma)$ as.
\begin{eqnarray}
M_{\nu} & =&
\left(
        \begin{array}{ccc}
         0  & A_{\nu} &  0 \\
        A_{\nu} & B_{\nu} & C_{\nu}\\
         0  & C_{\nu} & D_{\nu}
        \end{array}
\right) \nonumber \\ 
 & =& -s'/(r'^2 \gamma)
\left(
	\begin{array}{ccc}
		0 & A_D & 0 \\ 
		A_D & B_D & C_D \\
		0 & C_D & D_D 
	\end{array}
\right) 
\left(
		\begin{array}{ccc}
		0 & A_{S'} & 0 \\ 
		A_{S'} & B_{S'} & C_{S'} \\
		0 & C_{S'} & D_{S'} 
		\end{array}
	\right)^{-1} 
\left(
	\begin{array}{ccc}
		0 & A_D & 0 \\ 
		A_D & B_D & C_D \\
		0 & C_D & D_D 
	\end{array}
\right),\label{eq99061601}
\end{eqnarray}
\begin{eqnarray}
A_\nu &=& -\frac{s'}{r'^2 \gamma}
  \Bigg(\frac{{A_D^2}}{A_{S'}}\Bigg), \nonumber \\
B_\nu &=& -\frac{s'}{r'^2 \gamma}\Bigg\{\frac{A_D\ B_D}{A_{S'}}+C_D\ 
\Bigg(\frac{C_D}{D_{S'}}-\frac{A_D\ C_{S'}}{A_{S'}\ D_{S'}}\Bigg) \nonumber \\
&&+A_D\ \Bigg(\frac{B_D}{A_{S'}}-\frac{C_D\ C_{S'}}{A_{S'}\ D_{S'}}
-\frac{A_D\ \big(-{{C_{S'}}^2}+B_{S'}\ D_{S'}\big)}
{{{A_{S'}}^2}\ D_{S'}}\Bigg)\Bigg\},
\nonumber \\
C_\nu &=& -\frac{s'}{r'^2 \gamma}
 \Bigg\{\frac{A_D\ C_D}{A_{S'}}+D_D\ \Bigg(\frac{C_D}{D_{S'}}
     -\frac{A_D\ C_{S'}}{A_{S'}\ D_{S'}}\Bigg)\Bigg\}, \nonumber\\
D_\nu &=& -\frac{s'}{r'^2 \gamma}\Bigg(\frac{{{D_D}^2}}{D_{S'}}\Bigg).
\end{eqnarray}
Therefore, the neutrino mass ratios $m_1/m_2$ and 
$m_2/m_3$ and hence MNS 
lepton mixing matrix elements are also determined as functions of 
$\cos\alpha_{12}$ and $r$ .
The common overall factor $s'/(r'^2 \gamma)$ is 
determined by the $\Delta m^2$ data from neutrino oscillation 
experiments. 
The light Majorana neutrino masses are obtained by diagonalizing $M_{\nu}$ as 
\begin{equation}
O^T_{\nu}
\left(
        \begin{array}{ccc}
         0  & A_{\nu} &  0 \\
        A_{\nu} & B_{\nu} & C_{\nu}\\
         0  & C_{\nu} & D_{\nu}
        \end{array}
\right) 
O_{\nu}=
\left(
        \begin{array}{ccc}
        m_1 & 0 & 0\\
        0 & m_2 & 0\\
        0 &  0 & m_3
        \end{array}
\right).
\end{equation}
For the case in which $B_{\nu}, C_{\nu}, D_{\nu} \gg A_{\nu} $ is satisfied, the neutrino masses 
are approximately expressed in terms of $A_{\nu}$, $B_{\nu}$, $C_{\nu}$, and $D_{\nu}$ as 
\begin{eqnarray}
m_1 & \simeq& -\frac{D_{\nu}A^2_{\nu}}{B_{\nu}D_{\nu}- C^2_{\nu}} , \nonumber \\
m_2 & \simeq& \frac{1}{2}\{B_{\nu}+D_{\nu}-\sqrt{(B_{\nu}+D_{\nu})^2-4(B_{\nu}D_{\nu}- C^2_{\nu})} \}  \nonumber \\
	&   +& \frac{D^2_{\nu}+C^2_{\nu}-\frac{1}{2}D_{\nu}\{B_{\nu}+D_{\nu}-\sqrt{(B_{\nu}+D_{\nu})^2-4(B_{\nu}D_{\nu}- C^2_{\nu})} \}}
	{( B_{\nu}D_{\nu}- C^2_{\nu})\sqrt{(B_{\nu}+D_{\nu})^2-4(B_{\nu}D_{\nu}- C^2_{\nu})}}A^2_{\nu} , \\
m_3 & \simeq& \frac{1}{2}\{B_{\nu}+D_{\nu}+\sqrt{(B_{\nu}+D_{\nu})^2-4(B_{\nu}D_{\nu}- C^2_{\nu})} \}.\nonumber \\
	&   -& \frac{D^2_{\nu}+C^2_{\nu}-\frac{1}{2}D_{\nu}\{B_{\nu}+D_{\nu}+\sqrt{(B_{\nu}+D_{\nu})^2-4(B_{\nu}D_{\nu}- C^2_{\nu})} \}}
	{( B_{\nu}D_{\nu}- C^2_{\nu})\sqrt{(B_{\nu}+D_{\nu})^2-4(B_{\nu}D_{\nu}- C^2_{\nu})}}A^2_{\nu} . \nonumber 
\end{eqnarray}
The orthogonal matrices $O_{\nu}$ which diagonalizes $M_{\nu}$ 
are expressed in terms of the diagonalized masses $m_1$, $m_2$, and $m_3$ and 
the matrix components  $A_{\nu}$, $B_{\nu}$, $C_{\nu}$, and $D_{\nu}$ as \cite{kang2} 
\begin{equation}
O_{\nu}=
\left(
        \begin{array}{ccc}
         \frac{A_{\nu}}{m_1}(O_{\nu})_{21}    & \frac{A_{\nu}}{m_2}(O_{\nu})_{22} & \frac{A_{\nu}}{m_3}(O_{\nu})_{23} \\
        (O_{\nu})_{21} & (O_{\nu})_{22} & (O_{\nu})_{23}\\
         \frac{C_{\nu}}{m_1-D_{\nu}}(O_{\nu})_{21}    & \frac{C_{\nu}}{m_2-D_{\nu}}(O_{\nu})_{22} & \frac{C_{\nu}}{m_3-D_{\nu}}(O_{\nu})_{23}
        \end{array}
\right), \label{eq090711}
\end{equation}
with
\begin{eqnarray}
(O_{\nu})^2_{21} & =& \frac{1}{(\frac{A_{\nu}}{m_1})^2+1+(\frac{C_{\nu}}{m_1-D_{\nu}})^2} ,  \nonumber \\
(O_{\nu})^2_{22} & =& \frac{1}{(\frac{A_{\nu}}{m_2})^2+1+(\frac{C_{\nu}}{m_2-D_{\nu}})^2} , \label{eq090712} \\
(O_{\nu})^2_{23} & =& \frac{1}{(\frac{A_{\nu}}{m_3})^2+1+(\frac{C_{\nu}}{m_3-D_{\nu}})^2} .  \nonumber
\end{eqnarray}
It should be remarked that the light neutrino mass matrix \(M_{\nu}\) itself
is out of type I via the seesaw mechanism and that the MNS lepton mixing
matrix is obtained from Eqs. (\ref{eq090711}) and (\ref{123002}).  
Since $O_e$ is almost diagonal, the
magnitudes of off-diagonal elements are  predominated by Eq.(\ref{eq090712}). 
Thus the seesaw mechanism changes the form of lepton mixing 
matrix from that of CKM matrix given by Eq.(\ref{eq122003}).
\par
Now, by changing the values of the free parameters in our model,
we proceed to find the solutions 
which are consistent with the recent following findings that 
(i) the atmospheric neutrino oscillation experiment indicates
the \(\nu_\mu\)-\(\nu_\tau\) large mixing 
(\(0.28 \lsim |U_{23}| \lsim 0.72\) \cite{fukuyama}) with  
$\Delta m_{23}^2 = m^2_3-m^2_2 = (1.5\sim 6) \times10^{-3} \simeq 3.5\times10^{-3}\mbox{eV}^2$ , 
and (ii) the solar neutrino experiments imply the MSW small mixing angle  
solution\cite{bahcall} with 
$\Delta m_{12}^2 = m^2_2-m^2_1 = (4\sim 10)\times 10^{-6} \mbox{eV}^2$ and 
\(\sin^2 2\theta_{12}\) \(=\) \((2\sim 10)\times 10^{-3}\),
or 
the large mixing angle solution \cite{bahcall} with 
$\Delta m_{12}^2 = m^2_2-m^2_1\simeq (8\sim 30)\times 10^{-6} \mbox{eV}^2$ 
and 
\(\sin^2 2\theta_{12}\) \(=\) \((0.5 \sim 1)\).
In the following analysis, we transform $A_{e}$, $B_{e}$, $C_{e}$, 
and $D_{e}$ in Eq.(\ref{eq051101}) into $-A_{e}$, $-B_{e}$, $-C_{e}$, 
and $-D_{e}$ , respectively 
by rephasing of the right-handed charged lepton fields. 
\par
First assuming that the mass matrices 
$M_u$, $M_d$ and $M_e$ are all of type I, 
we calculate numerically the MNS lepton mixing matrix $U$ 
using the central values for the running quarks and 
charged leptons masses at \(\mu=m_Z\) \cite{Fusaoka}:
\begin{equation}
\begin{array}{lll}
m_u(m_Z)=2.33^{+0.42}_{-0.45}\mbox{MeV},& 
m_c(m_Z)=677^{+56}_{-61}\mbox{MeV},& 
m_t(m_Z)=181 \pm 13\mbox{GeV},\\
m_d(m_Z)=4.69^{+0.60}_{-0.66}\mbox{MeV},& 
m_s(m_Z)=93.4^{+11.8}_{-13.0}\mbox{MeV},& 
m_b(m_Z)=3.00 \pm 0.11\mbox{GeV},\\
m_e(m_Z)=0.487\mbox{MeV},& 
m_{\mu}(m_Z)=103\mbox{MeV},& 
m_{\tau}(m_Z)=1.75 \mbox{GeV}.
\end{array}
\label{eq123103}
\end{equation}
Since the recent 
atmospheric neutrino oscillation data indicates large value of (2,3) 
element of $U$, 
(\(0.28 \lsim |U_{23}| \lsim 0.72\) \cite{fukuyama}), we obtain 
the allowed 
region of the parameters space, $\cos\alpha_{12}$ vs $r$ space 
which reproduces 
a large $|U_{23}|$. The result is given in Fig. 1. In this allowed 
parameter region, 
$r \simeq 3$ is automatically satisfied without any fine tuning. 
However, we have a serious problem 
that in this allowed parameter space we cannot accommodate the overall 
factor  
$s'/(r'^2 \gamma)$ simultaneously to the data $\Delta m_{12}^2 = m^2_2-m^2_1\lsim 
10^{-4}\mbox{eV}^2$ (Here we have adopted a rather conservative value.
We accept more restrictive ones later.) from 
solar neutrino oscillation experiments and the data 
$\Delta m_{23}^2 = m^2_3-m^2_2\simeq 3.5\times10^{-3}\mbox{eV}^2$ 
from atmospheric 
neutrino
oscillation experiments. Taking deviations from the central values 
\cite{Fusaoka} for quarks and 
charged leptons masses does not resolve this problem. 
This difficulty is resolved by abandoning the above assumption 
that the mass matrices $M_u$, $M_d$ and $M_e$ are all of type I. 
Let us assume that $M_e$ deviates from type I 
although $M_u$ and $M_d$ are of type I. 
Then, we can accommodate the overall factor  
$s'/(r'^2 \gamma)$ simultaneously to both  $\Delta m^2$ data from 
solar and atmospheric neutrino oscillation experiments. 
\par
Next we discuss this new scenario and show that there are solutions 
consistent with the data.
First we represent the deviation from type I as 
\(B_e=m_\mu(1+\xi)\).  In this case, the entries of 
the mass matrix $M_e$ for charged leptons 
in Eq.(\ref{eq051101}) are given, in the unit of eV, as 
\begin{eqnarray}
&&A_e = \sqrt{\frac{m_e m_\mu m_\tau}{m_\tau-\xi\ m_\mu-m_e}} \simeq 7.1 \times 10^6, \nonumber \\
&&B_e =  (1+\xi) m_\mu \simeq 1.02 \times 10^8(1+\xi), \nonumber \\
&&C_e = \sqrt{(m_e+\xi\ m_\mu)\ (m_\tau-(1+\xi)\ m_\mu-m_e)\left(\frac{m_\tau-\xi\ m_\mu}{m_\tau-\xi\ m_\mu-m_e}\right)} \simeq 
  \sqrt{8.1 \times 10^{14}+1.7 \times 10^{17}\xi},
  \nonumber\\
&&D_e = -m_e-\xi\ m_\mu+m_\tau \simeq 7.1 \times 10^9,
\end{eqnarray}
and the expressions for \(\epsilon\), \(\alpha\) and the elements of the matrices $S$ and $S'$ are given by 
\begin{eqnarray}
\epsilon &\simeq& -\frac{(-r (1+\xi) m_\mu+3 m_s) m_t-m_c (3 m_b-r m_\tau)}
                  {r ((1+\xi) m_b m_\mu-m_s m_\tau)}, \nonumber \\
\alpha &\simeq& -\frac{r ((1+\xi) m_b m_\mu-(1+\xi) m_d m_\mu-m_s m_\tau)}
                {m_b m_c+(-r m_\mu-r \xi m_\mu-m_s) m_t-m_c (m_d-r m_\tau)}, 
                \nonumber \\
S_{12} &\simeq& \big((r m_\mu+m_s)
                  \big(\cos \alpha_{12} {\sqrt{m_d m_s}}
                   (-3 m_b m_c-r (1+\xi) m_\mu m_t)+ \nonumber \\
          &&      \cos \alpha_{12} {\sqrt{m_d m_s}} (3 m_s m_t+r m_c m_\tau)-
                r (-m_b m_\mu+m_s m_\tau) {\sqrt{m_c m_u}}\big)\big)\big/ 
                \nonumber \\
           &&      (4 r m_s (m_b m_\mu-m_s m_\tau)), \nonumber \\
S_{22}&\simeq& -\frac{(-r m_\mu-m_s) (-r (1+\xi) m_\mu+3 m_s) m_t}
                {4 r (m_b m_\mu-m_s m_\tau)}, \nonumber \\
S_{23}&\simeq& \big(-\cos \alpha_{23} {\sqrt{m_b m_d}} (-r m_\mu-m_s)m_t
                \nonumber \\
		&&		(-3 m_b m_c-r (1+\xi) m_\mu m_t+3 m_s m_t+r m_c m_\tau)- 
		        \nonumber \\
		&&		r (-r m_\mu-m_s) m_t ((1+\xi) m_b m_\mu-m_s m_\tau) 
                {\sqrt{m_t m_u}}\big)\big/ \nonumber \\
		 &&	\big({r^2} m_\tau (m_b m_c m_\mu+m_s (\xi m_\mu m_t-m_c m_\tau))- 
		       \nonumber \\
		&&	   r \big(3 {{m_b}^2} m_c m_\mu+4 {{m_s}^2} m_t m_\tau-  
               m_b m_s ((4+\xi) m_\mu m_t+3 m_c m_\tau)\big)\big), \nonumber \\
S_{33}&\simeq& -\frac{(-r m_\mu-m_s) m_t (3 m_b-r m_\tau)}
                {4 r (m_b m_\mu-m_s m_\tau)}, \nonumber \\
S'_{12}&\simeq& \frac{\big(\cos \alpha_{12} {\sqrt{m_d m_s}}
                 (r m_\mu+m_s) m_t-r (m_b m_\mu-m_s m_\tau)
                  {\sqrt{m_c m_u}}\big)}{4 m_s m_t}, \nonumber\\
S'_{22}&\simeq& \frac{1}{4} (r m_\mu+m_s), \nonumber \\
S'_{23}&\simeq& -\bigg(\cos \alpha_{23} {\sqrt{m_b m_d}}+
                 \frac{r (m_b m_\mu-m_s m_\tau) m_u}
                 {(-r m_\mu-m_s) {\sqrt{m_t m_u}}}\bigg)\Big/ \nonumber \\
             &&  \bigg(-1+\frac{(m_b m_\mu-m_s m_\tau)
                 ((-r (1+\xi) m_\mu+3 m_s) m_t- m_c (3 m_b-r m_\tau))}
                {(-r m_\mu-m_s) m_t ((1+\xi) m_b m_\mu-m_s m_\tau)}\bigg), 
                \nonumber\\
S'_{33}&\simeq& \frac{1}{4} (m_b-m_d+r m_\tau). \label{eq090702}
\end{eqnarray}
The point \(r\simeq 3\) is rather singular in the following sense.
As is seen from Eq.(\ref{eq090701}), 
\(\epsilon\) becomes small at \(r \simeq 3\), 
hence we can not neglect \(\xi\) in this region.
That is, \(\epsilon\) is sensitive to small \(\xi\).
For instance, substituting Eq.(\ref{eq090702}) into Eqs.(\ref{eq042201}) 
we obtain \(M_D\) and \(M_R\) and therefore \(M_\nu\) 
through Eq.(\ref{eq99061601}).
The behaviors of the elements in the neutrino mass matrix \(M_\nu\) are depicted in Fig.2.
The large \(\nu_\mu\)-\(\nu_\tau\) neutrino mixing appears 
under the condition that \(B_\nu \simeq D_\nu\) 
which is realized at \(\xi\simeq 0.02\).
By changing $\xi$ freely with fixed \(r(=3)\),  we can well 
reproduce the experimental data as shown 
in Figs. 3-5, in which the constraints from 
\(|U_{23}|\), \(\Delta m_{12}^2 / \Delta m_{23}^2 \), and 
the both are satisfied, respectively. 
It is seen from Fig. 5 that by deviating 
\(B_e\) a little bit from type I (\(\xi\sim0.01\)),
we can well reproduce the experimental data 
for the solar neutrino oscillation and 
atmospheric neutrino deficit.
If we relax the condition \(r=3\) and change 
\(r\) freely as well as \(\ca\) and 
\(\xi\) around the values of the above solutions, 
we have the larger allowed region  as is shown in Fig. 6.
In the above allowed regions shown in Figs. 1-5,
we have used only the conservative condition for \(\Delta m_{12}^2 \) from 
the solar neutrino experiments,  
that is, we have not used the constraints of the mixing angle from 
the solar neutrino oscillation experiments.  When we take them into 
account in addition to the constraints from \(\Delta m_{12}^2 \), 
we obtain more restrictive allowed region 
than that of Fig. 6.
Under the condition of the small mixing angle solution for solar neutrino experiments,
the larger region of \(|\ca|\) in Fig. 6 is eliminated 
and we have the allowed region  as is shown in Fig. 7.
On the other hand, under the condition of the large mixing angle solution, 
the smaller region of \(|\ca|\) is eliminated and
the allowed region is given in Fig. 8.
It should be noted that as seen in Figs. 6-8 
our model not only satisfies the experimental 
observations in the lepton sector 
but also provides the restriction  on the CP violation phase, \(\ca\), 
from the neutrino oscillation experiments. 
Of course, we can also restrict the other CP violation phases, 
\(\cos \alpha_{23}\), 
\(\cos \beta_{12}\) and \(\cos \beta_{23}\),  
which are respectively depicted in Fig.9-11. 
Also it follows from Eq. (\ref{eq99061601}) that the
neutrino mass ratios \(|m_1/m_2|\) and \(|m_2/m_3|\) 
become sensitive functions of \(\xi\), 
as are shown in Fig.12 taking typical values of \(r\) and \(\ca\).
The common overall factor $s'/(r'^2 \gamma)$ in Eq.(\ref{eq99061601}) is
determined to be of order $10^{-13}$  by the $\Delta m^2$ data  from  the
solar and atmospheric neutrino oscillation experiments.
\par
Finally we discuss the entries of the CKM quark mixing matrix which are given 
by Eqs.(\ref{eq122003}) with (\ref{eq04190}). In our model, 
all the elements of the CKM mixing matrix are 
also functions of two free parameters \(\ca\) and \(\xi\). 
The parameters determined so far from the lepton sector do not give rise any 
inconsistency with the data in quark sector.
\par
\section{summary}
In this paper we have presented and discussed a model of  
texture four zero quark-lepton mass matrices 
in the context of SO(10).  The consistent fitting of the free parameters 
to the data for neutrino oscillation experiments has 
forced us to use the charged lepton mass matrix 
which slightly deviates from purely type I form (\(\xi\sim0.01\)). 
Using this deviated type of mass matrix for the charged leptons and 
the mass matrices for quarks of type I,   
we have been able to reproduce four entries in the CKM quark mixing matrix and 
to predict six entries in the MNS lepton mixing matrix and 
three Majorana neutrino masses which are consistent with the experimental data.
The model has also given the restrictions on the $CP$ violating phases 
which came from the neutrino oscillation experiments. 
Remarkably enough the parameter \(r\) fixed from data fitting 
is coincident with the value \(r \simeq (2\sim 3)\) obtained 
from the renormalization equation \cite{buras}. 
So it is attractive to expect 
that the above deviation \((\xi\sim 0.01)\) from type I form 
can be obtained by taking the evolution equation of
Yukawa coupling fully.
(In this paper we have considered the loop correction 
of gauge boson in the evolution equation.)
Though the detail calculations will be developed in the forthcoming paper,
we will roughly outline our idea. That is, (charged lepton) mass matrix is 
exactly of type I at some scale.
However, they change their form due to the evolutionary equation 
of the Yukawa coupling \(Y_a\) until the corresponding Higgs field acquires
the vacuum expectation value \cite{Fusaoka}
\begin{equation}
\frac{dY_a}{dt}= \frac{1}{16\pi^2}(T^f-G^f+H^f)
\end{equation}
where \(T^f\), \(G^f\), \(H^f\) are the vertex corrections due to 
the fermion, the gauge boson and the Higgs boson, respectively. 
After that, each mass furthermore changes its value according to the
mass renormalization equation.
The evolution equation of Yukawa coupling is very sensitive to the Higgs 
potentials and the initial conditions. One such sensitivity has been 
found in the behavior of \(\xi\).  The detail will be given in the 
forthcoming paper. 

\ \\
Acknowledgements

We are greatly indebted to Y. Koide for many useful comments 
on the whole subject discussed in this article. 
We also thank the anonymous referee for very useful comments.

\begin{figure}
\begin{center}
 	\leavevmode
 	\epsfile{file=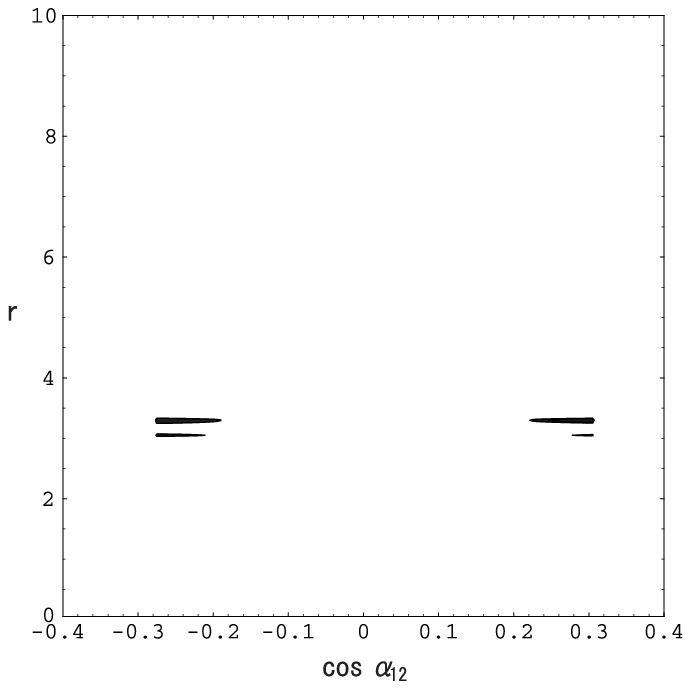}
\end{center}
\caption{
The case where \(M_u\), \(M_d\), and \(M_e\) are all of purely type I 
is analyzed.
The experimental constraint on $|U_{23}|$ (
\(0.28 \le |U_{23}|^2 \le 0.72\) )  gives the allowed region (shaded area) 
in the \(\ca\)-r plane. 
Here the \(r\) is treated as a free parameter.
}
\label{fig1}
\end{figure}

\begin{figure}
\begin{center}
 	\leavevmode
 	\epsfile{file=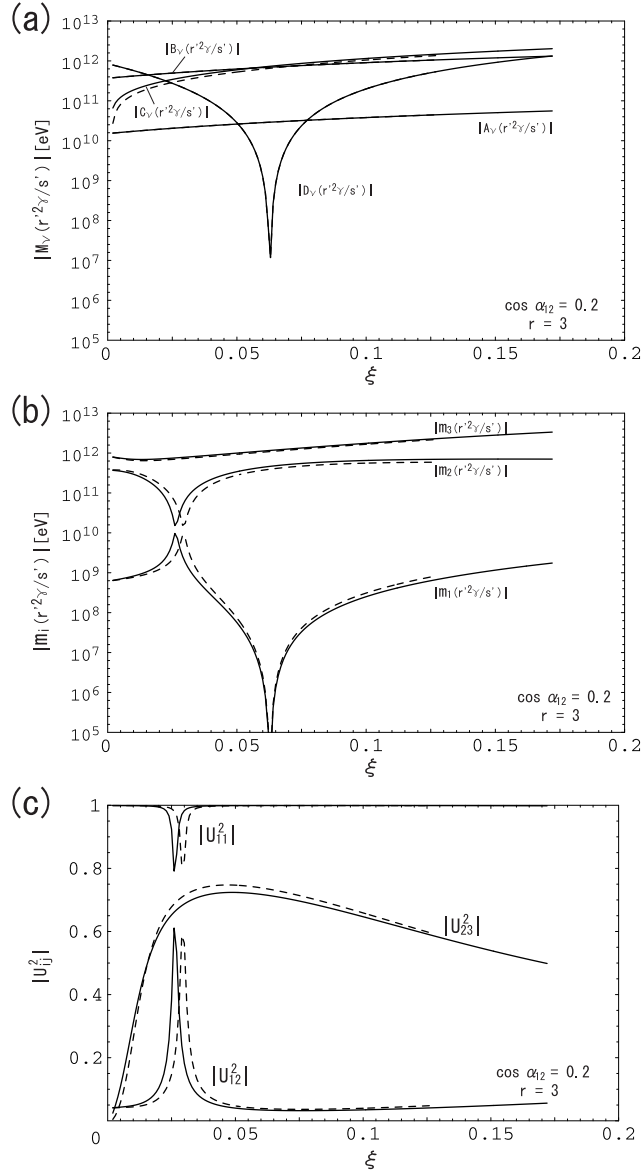}
\end{center}
\caption{%
The slight deviation from type I \((\xi \neq 0)\) makes 
physical parameters change drastically .
The dotted lines (solid lines) show the \(\xi\) dependence for 
\(\cos \alpha_{23} \ge 0\) 
(\(\cos \alpha_{23}\le 0\)) in each diagram.
All lines terminate at the points from where 
\(|\cos\alpha_{23}|\ge 1\) or 
\(|\cos \beta_{23}|\ge 1\) as will be seen from Fig.9 and Fig.11.
(a) The diagram of the elements in the neutrino mass matrix \(M_\nu\) 
versus \(\xi\).
Except for \(|C_\nu(r'^2 \gamma/s')|\) 
the dotted lines are overlapped with the corresponding solid lines.
(b) The diagram of the neutrino mass eigen values versus \(\xi\).
(c) The MNS mixing matrices versus \(\xi\).
}
\label{fig2}
\end{figure}

\begin{figure}
\begin{center}
 	\leavevmode
 	\epsfile{file=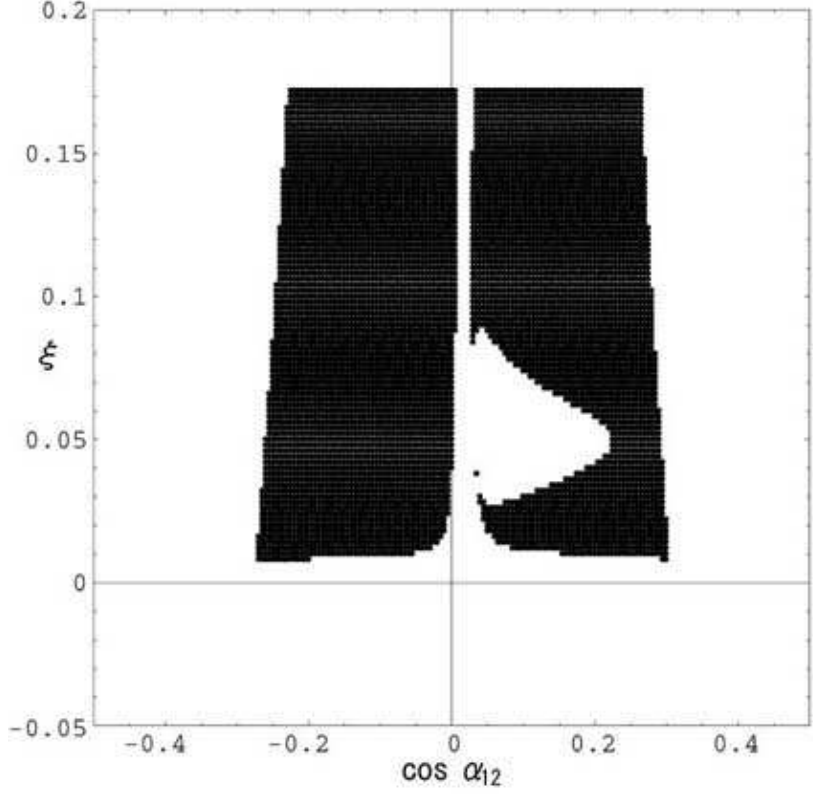}
\end{center}
\caption{
The experimental constraint on $|U_{23}|$ (
\(0.28 \le |U_{23}|^2 \le 0.72\) ) 
gives the allowed region (dotted area) 
in the \(\ca\)-\(\xi\) plane. }
\label{fig3}
\end{figure}

\begin{figure}
\begin{center}
 	\leavevmode
 	\epsfile{file=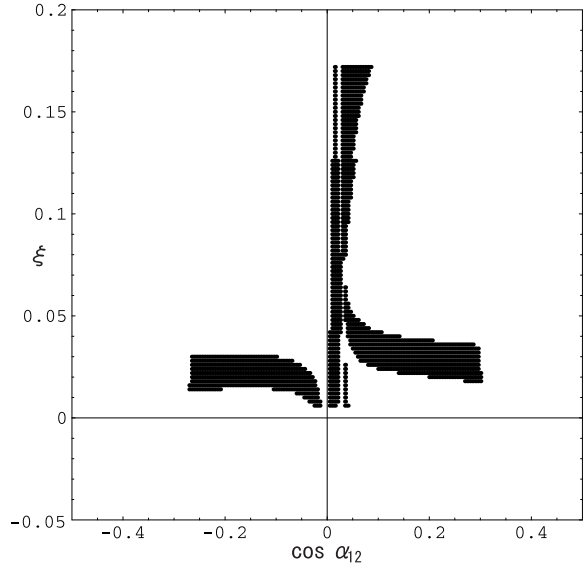}
\end{center}
\caption{
The allowed region in the 
\(\cos \alpha_{12}\) - \(\xi\) plane from
the experimental constraints 
\(\Delta m^2_{12}/\Delta m^2_{23}\) \(\le\) 
\((1\times 10^{-4})/(3.5 \times 10^{-3})\) \(=\) 
\(2.9 \times 10^{-2}\).}
\label{fig4}
\end{figure}

\begin{figure}
\begin{center}
 	\leavevmode
 	\epsfile{file=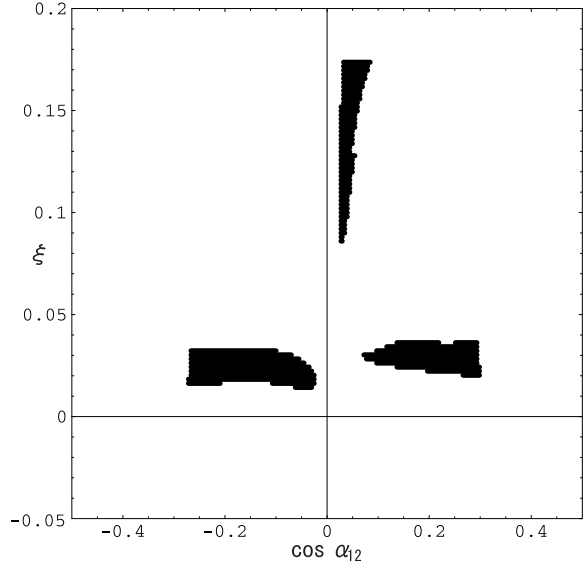}
\end{center}
\caption{
The allowed region in the 
\(\cos \alpha_{12}\) - \(\xi\) plane from
the experimental constraints \(0.28 \le |U_{23}|^2 \le 0.72\) and 
\(\Delta m^2_{12}/\Delta m^2_{23}\) \(\le\) \(2.9 \times 10^{-2}\).
}
\label{fig5}
\end{figure}

\begin{figure}
\begin{center}
 	\leavevmode
 	\epsfile{file=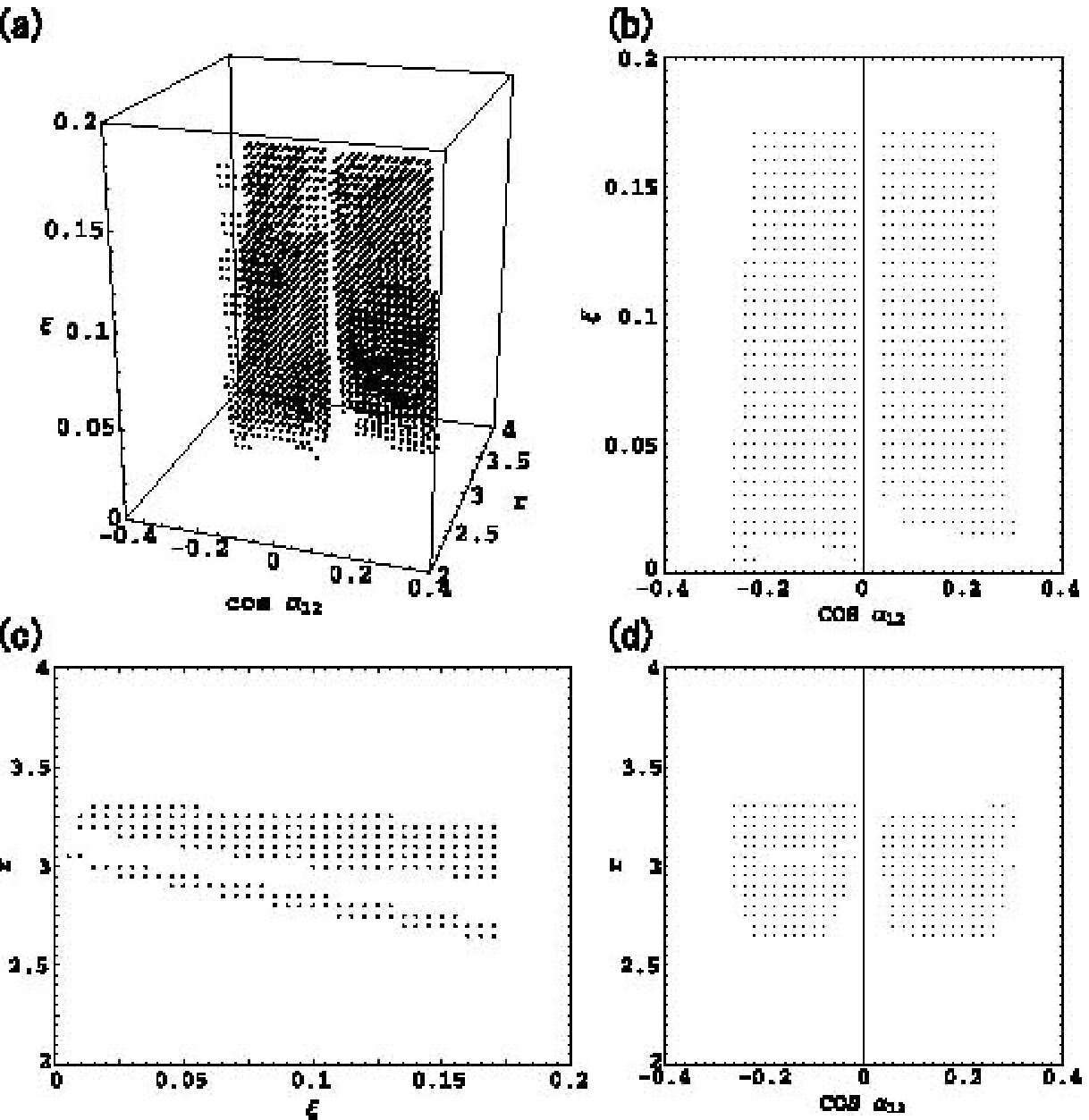}
\end{center}
\caption{
The \(r\) is treated as a free parameter.
(a) shows the allowed region in the 
\(\cos \alpha_{12}\) - \(r\) - \(\xi\) space 
from the experimental constraints  \(0.28 \le |U_{23}|^2 \le 0.72\) and 
\(\Delta m^2_{12}/\Delta m^2_{23}\) \(\le\) \(2.9 \times 10^{-2}\).
(b), (c) and (d) show 
the projected allowed regions in the \(\ca\)-\(\xi\), \(\xi\)-\(r\), and 
\(\ca\)-\(r\) planes, respectively.
}
\label{fig6}
\end{figure}

\begin{figure}
\begin{center}
 	\leavevmode
 	\epsfile{file=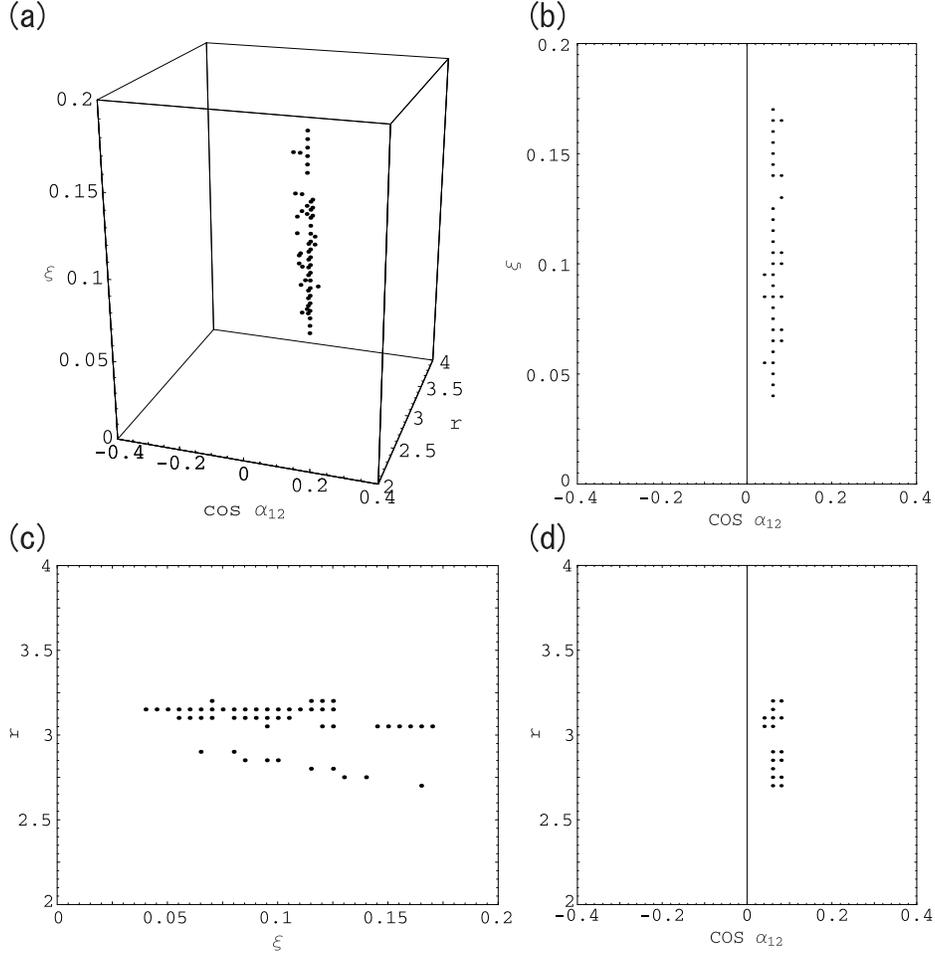}
\end{center}
\caption{
The allowed region in  the \(\ca\)-\(r\)-\(\xi\) space
from the experimental constraints \(0.28 \le |U_{23}|^2 \le 0.72\), 
the small mixing angle solution of the solar neutrino experiments 
(\(\sin^2 2\theta_{12}\) \(=\) \((2\sim 10)\times 10^{-3}\)), 
and the up-to-date value of mass difference 
\(\Delta m^2_{12}/\Delta m^2_{23}\) \(=\) 
\(((4\sim 10)\times 10^{-6})/((1.5\sim 6) \times10^{-3})\) \(=\) 
\((0.67\sim 6.7)\times 10^{-3}\).
}
\label{fig7}
\end{figure}

\begin{figure}
\begin{center}
 	\leavevmode
 	\epsfile{file=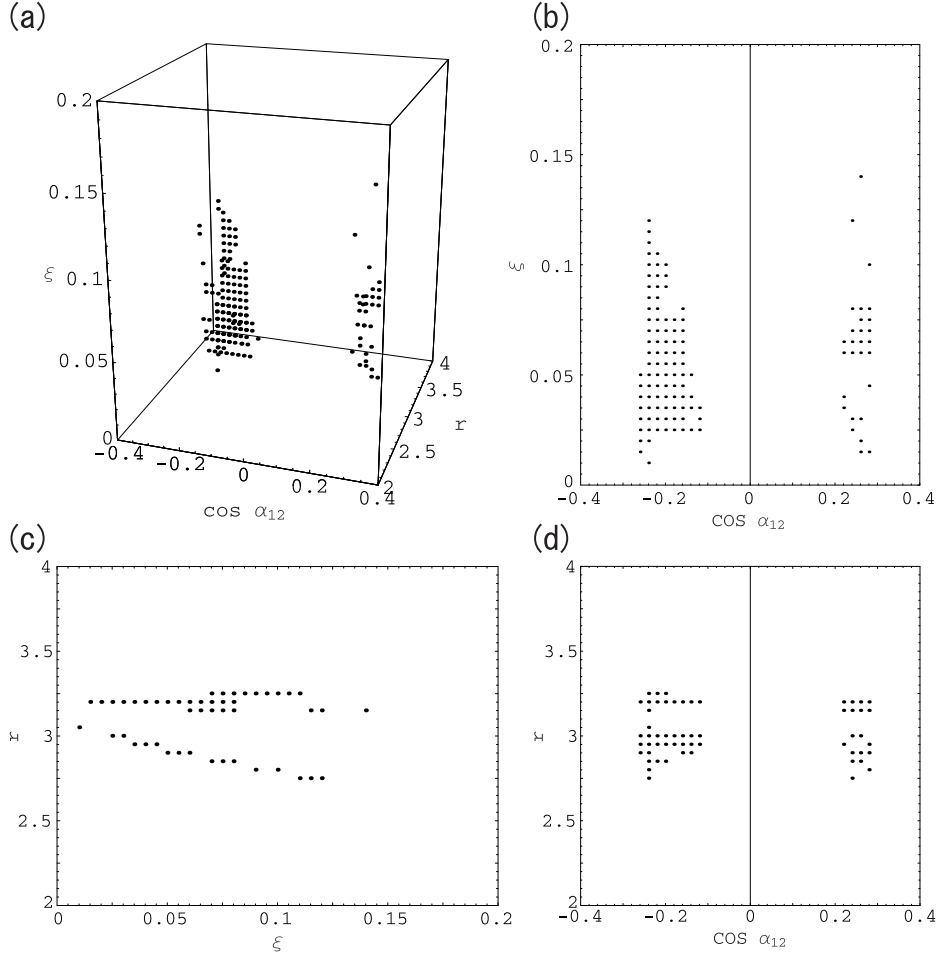}
\end{center}
\caption{
The allowed region in  the \(\ca\)-\(r\)-\(\xi\) space
from the experimental constraints \(0.28 \le |U_{23}|^2 \le 0.72\), 
the large mixing angle solution of the solar neutrino experiments 
\(\sin^2 2\theta_{12}\) \(=\) \((0.5 \sim 1)\),
and the up-to-date value of mass difference 
\(\Delta m^2_{12}/\Delta m^2_{23}\) \(=\) 
\(((8\sim 30)\times 10^{-6})/((1.5\sim 6) \times10^{-3})\) \(=\) 
\((0.13\sim 2.0)\times 10^{-3}\). 
}
\label{fig8}
\end{figure}

\begin{figure}
\begin{center}
 	\leavevmode
 	\epsfile{file=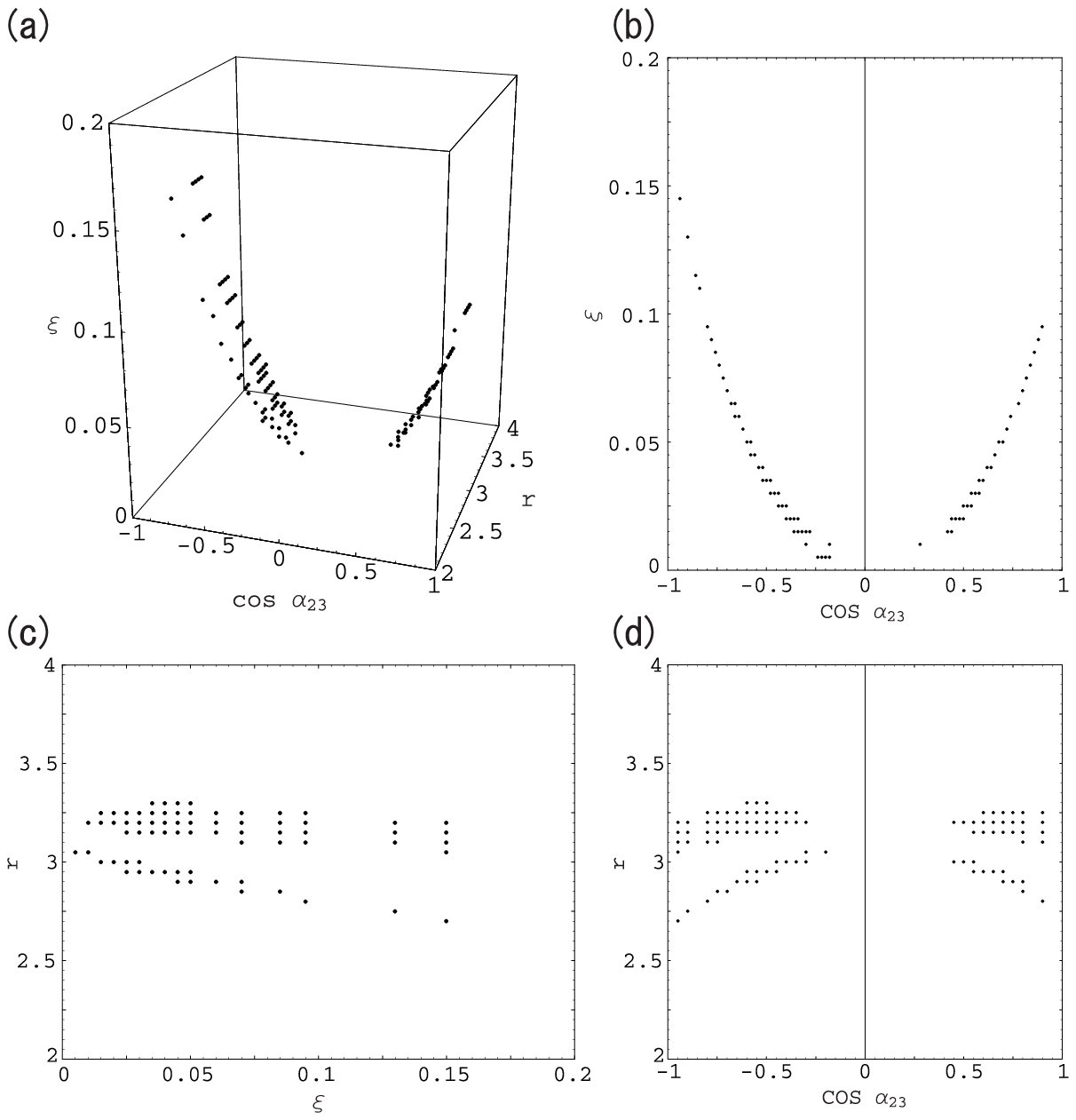}
\end{center}
\caption{
The allowed region in the 
\(\cos \alpha_{23}\) - \(r\) - \(\xi\) space from the same 
constraints as in Fig.6.
}
\label{fig9}
\end{figure}

\begin{figure}
\begin{center}
 	\leavevmode
 	\epsfile{file=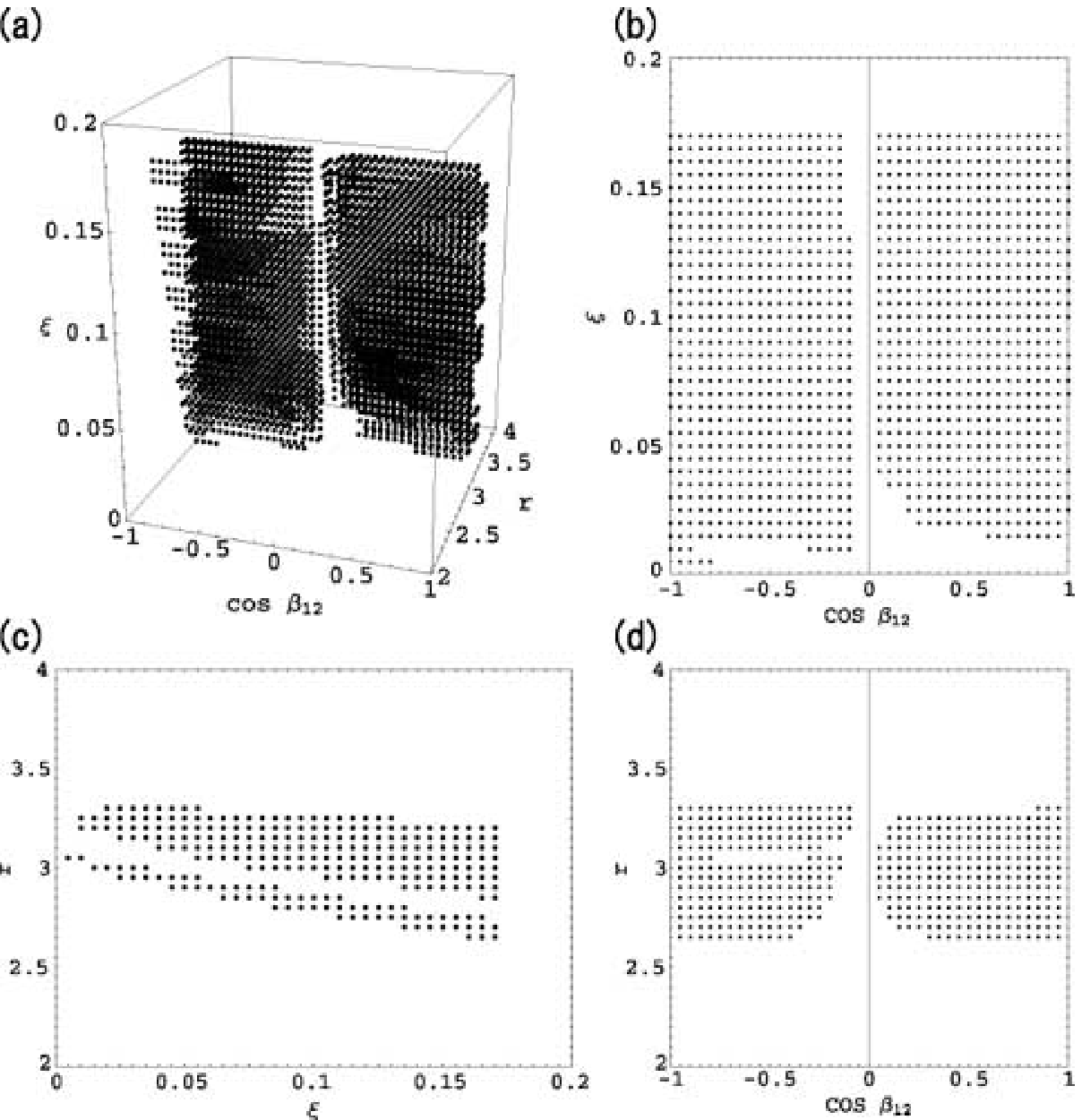}
\end{center}
\caption{
The allowed region in the 
\(\cos \beta_{12}\) - \(r\) - \(\xi\) space from the same 
constraints as in Fig.6.
}
\label{fig10}
\end{figure}

\begin{figure}
\begin{center}
 	\leavevmode
 	\epsfile{file=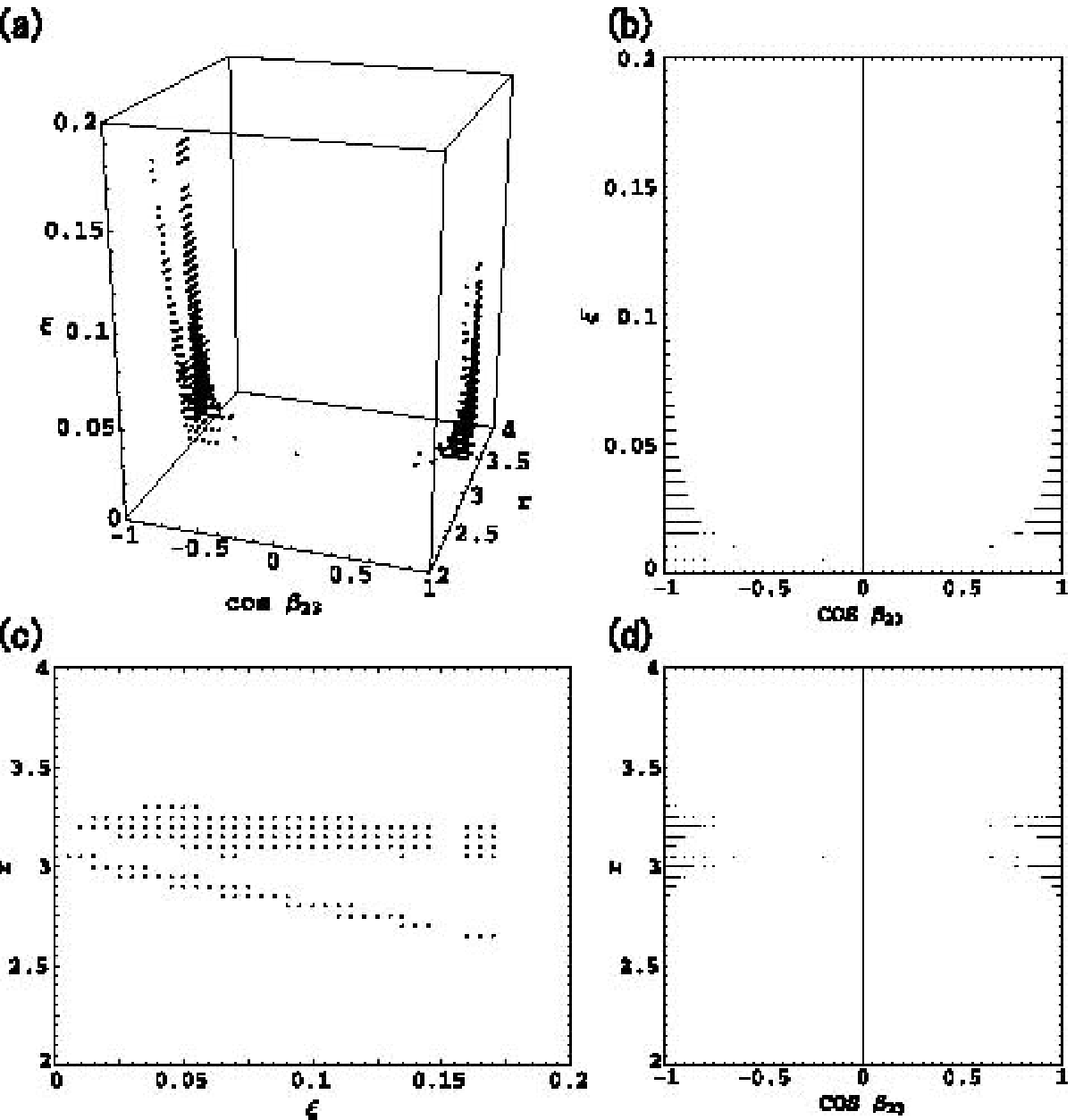}
\end{center}
\caption{
The allowed region in the 
\(\cos \beta_{23}\) - \(r\) - \(\xi\) space from the same 
constraints as in Fig.6.
}
\label{fig11}
\end{figure}

\begin{figure}
\begin{center}
 	\leavevmode
 	\epsfile{file=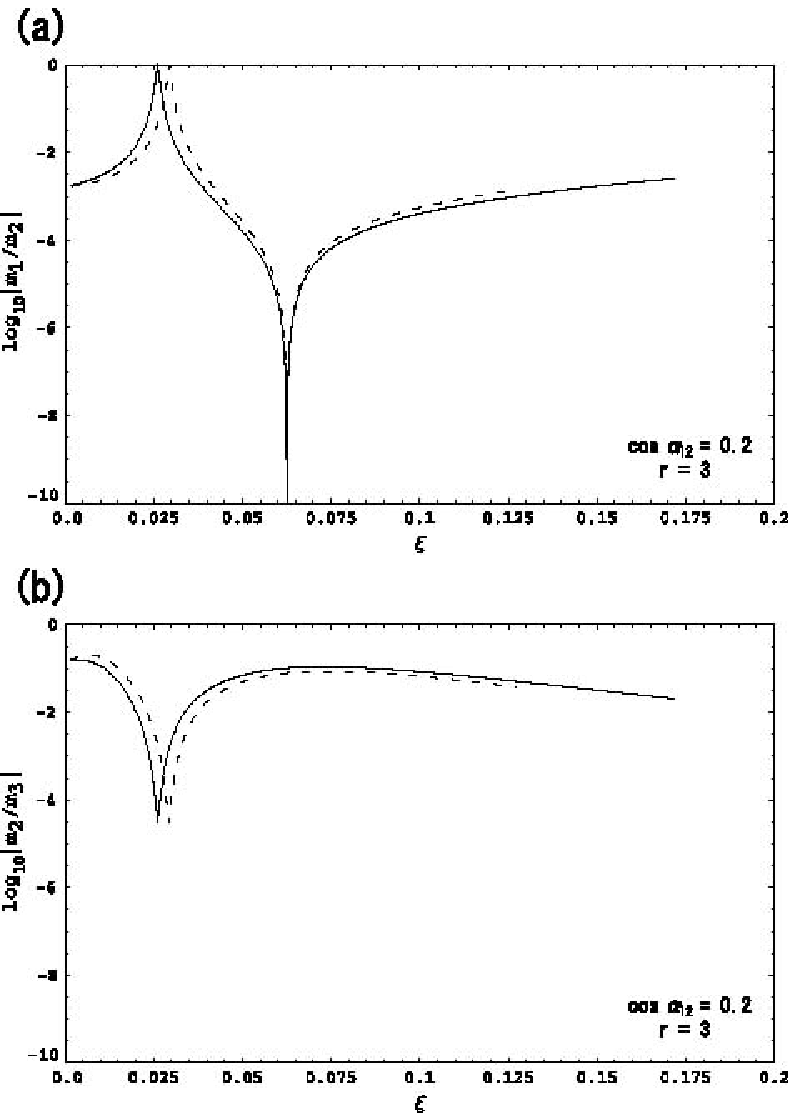}
\end{center}
\caption{
The dependence of \(\log_{10}|m_1/m_2|\) (a) and 
\(\log_{10}|m_2/m_3|\) (b) on \(\xi\) for \(\ca=0.2\) and \(r=3\).
The dotted line (solid line) shows the \(\xi\) dependence for 
\(\cos\alpha_{23} \ge 0\) (\(\cos\alpha_{23} \le 0\)) in each diagram.
Both lines terminate at the points from where \(|\cos\alpha_{23}|\ge 1\) or 
\(|\cos \beta_{23}|\ge 1\) as are seen from Fig.9 and Fig.11.
The singular behaviors of (a) and (b) come from those of \(m_1\) and \(m_2\)
(see Fig.2).}
\label{fig12}
\end{figure}


\begin{thebibliography}{99}
\bibitem{skamioka}
T. Kajita, Talk at Neutrino '98 (Takayama, Japan, June 1998); 
Y. Fukuda et. al., Phys. Rev. Lett. {\bf 81}, 1562 (1998).
\bibitem{skamioka2}
Y. Takeuchi, Observation of solar neutrinos in Super-Kamiokande-present 
and future- (Talk in Future od Neutrino Physics at KEK, Japan, March 
1999).
\bibitem{chooz}
M. Apollonio et.al., Phys. Lett. {\bf B420}, 397 (1998).
\bibitem{chorus}
K. Winter, Nucl. Phys. {\bf B38} (Proc. Suppl), 221 (1995)
\bibitem{yanagida}
T. Yanagida in Proceedings of the Workshop on the unified theory 
and baryon number in the 
Universe, 
edited by O. Sawada and A. Sugamoto (KEK, Tsukuba) (1979),p95;
M. Gell-Mann, P. Ramond and R. Slansky in  
Supergravity, edited by  P.van Nieuwenhuizen and D. Freedman 
(Amsterdam, North Holland,1979)p315; 
R.N. Mohapatora and G. Senjanovic, Phys. Rev. Lett. {\bf 44}, 912 (1980).
\bibitem{fritzsch}
H. Fritzsch, Nucl. Phys. {\bf B155}, 189 (1979).
\bibitem{ramond}
P. Ramond, R.G. Roberts, and G.G. Ross, Nucl. Phys. {\bf B406}, 19 
(1993).
\bibitem{branco}
G.C. Branco, L. Lavoura and F. Mota, Phys. Rev. {\bf D39}, 3443 (1989).
\bibitem{takasugi}
E. Takasugi, Prog. Theor. Phys. {\bf 98}, 177 (1997).
\bibitem{nishiura3}
H. Nishiura, K. Matsuda and T. Fukuyama, "Lepton and Quark Mass 
Matrices" to appear in Phys. Rev. D (hep-ph/9902385)
\bibitem{du}
D. Du and Z.Z. Xing, Phys. Rev. {\bf D48}, 2349 (1993).
\bibitem{fritzsch2}
H. Fritzsch and Z.Z. Xing, Phys. Lett. {\bf B353}, 114 (1995).
\bibitem{kang}
K. Kang and S.K. Kang, Phys. Rev. {\bf D56}, 1511 (1997).
\bibitem{kang2}
K. Kang, S.K. Kang, C.S. Kim and S.M. Kim, hep-ph/9808419.
\bibitem{chkareuli}
J.L. Chkareuli and C.D. Froggatt, Phys. Lett. {\bf B450}, 158 (1999).
\bibitem{shizuoka}
See, for example, Proceeding of the International Workshop on Masses and 
Mixings of Quarks and Leptons,  Shizuoka, Japan, March 1997 
edited by Y.  Koide ( World Scientific Publishing Co.,  1998).
\bibitem{MNS}
Z. Maki, M. Nakagawa and S. Sakata, Prog. Theor. Phys. {\bf 28}, 247 (1962).
\bibitem{bottino}
A. Bottino, C. W. Kim, H. Nishiura, and W. K. Sze, Phys. Rev. {\bf D34}, 862 
(1986). 
\bibitem{johnson}
R. Johnson, S. Ranfone, and J. Schechter, Phys. Rev. {\bf D35}, 282 (1987). 
\bibitem{bottino2}
A. Bottino, C. W. Kim, and H. Nishiura, Phys. Rev. {\bf D30}, 1046 
(1984). In this paper, the contribution of 126 is neglected in $M_u$.
\bibitem{chanowitz}
M.S. Chanowitz, J.E. Ellis and M.K. Gaillard, Nucl. Phys. {\bf B128}, 506 
(1977).
\bibitem{Harvey}
J.A. Harvey, D.B. Reiss, and P. Ramond, Nucl. Phys. {\bf B199}, 223 
(1982).
\bibitem{Vergados}
J.D. Vergados, Phys. Rep. {\bf 133} (1986).
\bibitem{buras}
A.J. Buras, J. Ellis, M.K. Gaillard and D.V. Nanopoulos, 
Nucl. Phys. {\bf B135}, 66 (1978). 
\bibitem{barbieri}
R. Barbieri, D.V. Nanopoulos, G. Morchio, and F. Strocchi, Phys. Lett. {\bf B90}, 91 (1982).
\bibitem{fukuyama}
T. Fukuyama, K. Matsuda, and H. Nishiura, 
Mod. Phys. Lett. {\bf A13}, 2279 (1998).
\bibitem{bahcall}
J.N. Bahcall, P.I. Krastev and A.Yu. Smirnov, 
Phys. Rev. {\bf D58}, 096016 (1998). 
\bibitem{Fusaoka}
H. Fusaoka and Y. Koide, Phys. Rev. {\bf D57}, 3986 (1998).

\end{thebibliography}
\end{document}